\documentclass[useAMS,usenatbib,usegraphicx,onecolumn]{mn2e}

\usepackage{amssymb,amsfonts,amsmath,amstext,amsgen,amsopn,amsxtra,indentfirst,lscape,times,rotating}

\title[Clouds and hazes in hot Jovian atmospheres]{On the effects of clouds and hazes in the atmospheres of hot Jupiters: semi-analytical temperature-pressure profiles}
\author[Heng, Hayek, Pont \& Sing (HHPS)]{Kevin Heng$^{1}$\thanks{E-mail: kheng@phys.ethz.ch, heng@ias.edu (KH)}, Wolfgang Hayek$^{2}$\thanks{Email: hayek@astro.ex.ac.uk  (WH)}, Fr\'{e}d\'{e}ric Pont$^{2}$\thanks{Email: fpont@astro.ex.ac.uk (FP)} and David K. Sing$^{2}$\thanks{Email: sing@astro.ex.ac.uk (DS)}\\
$^{1}$Zwicky Fellow, ETH Z\"{u}rich, Institute for Astronomy, Wolfgang-Pauli-Strasse 27, CH-8093, Z\"{u}rich, Switzerland\\
$^{2}$University of Exeter, School of Physics, Astrophysics Group, Stocker Road, Exeter EX4 4QL, United Kingdom}

\begin{document}

\date{Submitted 2011 July 6.  Re-submitted 2011 September 8.  Accepted 2011 October 3.}

\pagerange{\pageref{firstpage}--\pageref{lastpage}} \pubyear{2011}

\maketitle

\label{firstpage}

\begin{abstract}
Motivated by the work of Guillot (2010), we present a semi-analytical formalism for calculating the temperature-pressure profiles in hot Jovian atmospheres which includes the effects of clouds/hazes and collision-induced absorption.  Using the dual-band approximation, we assume that stellar irradiation and thermal emission from the hot Jupiter occur at distinct wavelengths (``shortwave" versus ``longwave").  For a purely absorbing cloud/haze, we demonstrate its dual effect of cooling and warming the upper and lower atmosphere, respectively, which modifies, in a non-trivial manner, the condition for whether a temperature inversion is present in the upper atmosphere.  The warming effect becomes more pronounced as the cloud/haze deck resides at greater depths.  If it sits below the shortwave photosphere, the warming effect becomes either more subdued or ceases altogether.  If shortwave scattering is present, its dual effect is to warm and cool the upper and lower atmosphere, respectively, thus counteracting the effects of enhanced longwave absorption by the cloud/haze.  We make a tentative comparison of a 4-parameter model to the temperature-pressure data points inferred from the observations of HD 189733b and estimate that its Bond albedo is approximately 10\%.  Besides their utility in developing physical intuition, our semi-analytical models are a guide for the parameter space exploration of hot Jovian atmospheres via three-dimensional simulations of atmospheric circulation.
\end{abstract}

\begin{keywords}
planets and satellites: atmospheres -- radiative transfer
\end{keywords}

\section{Introduction}
\label{sect:intro}

The study of extrasolar planets has swiftly transitioned from discovery to characterization.  Following the pioneering detection of the atmosphere of a hot Jupiter by \cite{char02}, researchers have used the transit method to infer the presence of clouds and hazes in these atmospheres (e.g., \citealt{pont08,sing11}).  (See \citealt{fortney10} and references therein for a detailed comparison of the observed atmospheric properties of hot Jupiters to spectral models.)  Such discoveries motivate the improvement of atmospheric models to include the effects of clouds and hazes.  In particular, the simplicity of one- or two-dimensional models allows one to develop physical intuition prior to including new physical effects in (expensive) three-dimensional simulations of atmospheric circulation.  The main purpose of the present study is to generalize the formalism presented in \cite{hubeny03}, \cite{hansen08} and \cite{guillot10} to include several new effects: scattering, collision-induced absorption and additional sources of longwave absorption.  The joint consideration of some of these improvements allows us to understand the effects of clouds/hazes on the temperature-pressure profiles of hot Jovian atmospheres.  The simplicity and versatility of our semi-analytical models allow for an easy comparison to present and future observations aiming to characterize the atmospheres of hot Jupiters.

Absorption and scattering introduce competing effects.  Enhanced longwave absorption by the cloud/haze has the \textit{dual} effect of cooling and warming the upper and lower atmosphere, respectively.  By contrast, if the cloud/haze scatters in the shortwave, it has the counteracting, dual effect of respectively warming and cooling the upper and lower atmosphere.  All of these effects combine to make the generalization of the condition for the presence or absence of a temperature inversion difficult, as opposed to its crisp one-parameter description in a cloud-free scenario \citep{hubeny03,hansen08,guillot10}.  As a demonstration of the utility of our models, we tentatively compare our semi-analytical, global-mean temperature-pressure profiles to the data points inferred from the observations of HD 189733b and estimate that its Bond albedo is about 0.1 (Figure \ref{fig:cloud_fit}).

In \S\ref{sect:formalism}, we describe the derivation of a general equation for the temperature-pressure profile in a hot Jupiter atmosphere, which allows for an arbitrary functional form for the longwave opacity $\kappa_{\rm L}$.  In \S\ref{sect:applications}, we derive more specific forms of the equation assuming various functional forms for $\kappa_{\rm L}$.  Examples of temperature-pressure profiles are calculated in \S\ref{sect:examples}.  We discuss the implications of our results in \S\ref{sect:discussion}.  Table \ref{tab:symbols} contains a list of the commonly used symbols in this study.  Our main technical results are stated in equations (\ref{eq:t4_global_cloud0}) and (\ref{eq:t4_global_cloud}).  Appendices \ref{append:two_stream}, \ref{append:deposition}, \ref{append:identity} and \ref{append:operators} archive technical information relevant to the formalism presented in \S\ref{sect:formalism}. 

\section{General formalism for temperature-pressure profile}
\label{sect:formalism}

\begin{figure}
\begin{center}
\includegraphics[width=0.5\columnwidth]{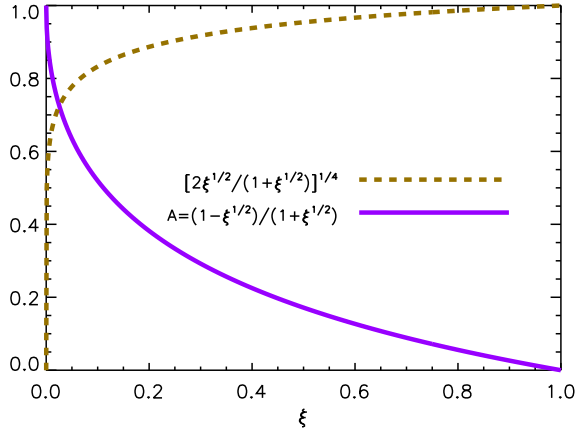}
\end{center}
\vspace{-0.2in}
\caption{Dependence of the Bond albedo ${\cal A}$ and the coefficient in $T_{\rm irr}$ (or $T_{\rm eq}$), $( 1 - {\cal A})^{1/4}$, on the scattering parameter $\xi$.}
\label{fig:albedo_xi}
\end{figure}

In this section, we aim to analytically derive a general expression for the global-mean temperature-pressure profile of a hot Jupiter atmosphere by generalizing the formalism presented in Section 3.2 of \cite{guillot10}.  We assume that stellar irradiation (``shortwave"; denoted by ``S") and thermal emission (``longwave"; denoted by ``L") from the hot Jupiter peak at distinct wavelengths, i.e., the dual-band approximation.  The zeroth, first and second moments of the specific intensity are represented by $J_\nu$, $H_\nu$ and $K_\nu$, respectively, where $\nu$ denotes the photon frequency.  The frequency-dependent absorption opacity is denoted by $\kappa_\nu$.  In general, the functional dependence of a given quantity is suppressed unless it is necessary to distinguish between functions with different arguments (e.g., equation [\ref{eq:h}]).

We consider both the effects of absorption and scattering, but treat them in an approximate manner in order to ensure algebraic tractability.  For simplicity, we include scattering only in the shortwave via the parameter $\xi$, which is the ratio of the absorption to the total opacity, while being aware that longwave scattering is a non-negligible effect \citep{dekok11}.\footnote{The formalism remains algebraically tractable even if the longwave analogue of $\xi$ is included as an additional parameter.}  Thus, we have $\xi=0$ and 1 in purely scattering and absorbing situations, respectively.  For example, \cite{guillot10} assumes $\xi=1$.  It is important to note that decreasing $\xi$ has the effect of increasing the number of shortwave scatterers present, while keeping the number of shortwave absorbers constant.  The Bond albedo ${\cal A}$ and $\xi$ are not independent parameters.  Rather, as a generic statement, we expect: when $\xi=0$, ${\cal A}=1$; conversely, when $\xi=1$, ${\cal A}=0$.  These properties serve as boundary conditions to the functional form $\xi = \xi({\cal A})$, which we will elucidate later, in the context of the collimated beam approximation (Appendix \ref{append:two_stream}), in equation (\ref{eq:albedo_xi}) and Figure (\ref{fig:albedo_xi}).

The shortwave optical depth is
\begin{equation}
\tau_{\rm S} = \int^m_0 \frac{\kappa_{\rm S}}{\xi} ~dm^\prime = \frac{\kappa_{\rm S} m}{\xi},
\end{equation}
where $m$ is the column mass per unit area and $\kappa_{\rm S}$ is the (constant) shortwave absorption opacity which we will describe shortly.  In a hydrostatically balanced atmosphere, the shortwave photosphere resides at
\begin{equation}
P_{\rm S} = \frac{\varpi g \xi}{\kappa_{\rm S}},
\label{eq:pressure_s}
\end{equation}
with $g$ being the surface gravity of the hot Jupiter.  The preceding expression follows from demanding that $\tau_{\rm S} = \varpi$.  In the solution to the classical Milne's problem for a self-radiating (stellar) atmosphere (see Section 3.3 of \citealt{mihalas}), one obtains $\varpi = 2/3 \approx 0.67$.  For irradiated atmospheres, the value of $\varpi$ needs to be estimated numerically (see Appendix \ref{append:deposition}); surprisingly, its value is not very different: $\varpi \approx 0.63$.  The \textit{photon deposition depth} --- the location where the energy of the stellar photons are deposited --- resides slightly deeper than the shortwave photosphere: at $\sqrt{\xi} \tau_{\rm S} = \varpi$ instead of $\tau_{\rm S}=\varpi$ (see equation [\ref{eq:shortwave3}]).  Thus, the pressure level at which shortwave photon deposition occurs is
\begin{equation}
P_{\rm D} = \frac{\varpi g \sqrt{\xi}}{\kappa_{\rm S}} = \frac{P_{\rm S}}{\sqrt{\xi}} = 63 \mbox{ mbar} ~\sqrt{\xi} \left( \frac{\varpi}{0.63} \frac{g}{10 \mbox{ m s}^{-2}} \right) \left( \frac{\kappa_{\rm S}}{0.01 \mbox{ cm}^2 \mbox{ g}^{-1}} \right).
\label{eq:deposition}
\end{equation}
As the effect of scattering becomes stronger (decreasing $\xi$), the altitude at which the stellar photons are absorbed becomes higher if shortwave absorption stays constant.  In the extreme limit of $\xi=0$, no starlight gets absorbed and the atmosphere has an albedo of unity --- its temperature-pressure profile is described only by its internal heat flux.  Since $\xi \le 1$, we have $P_{\rm D} \ge P_{\rm S}$.  When shortwave scattering is absent, we obtain $P_{\rm S}=P_{\rm D}$.

For the rest of the paper, we refer only to \textit{absorption} (and not scattering) opacities when we use the term ``opacity" (unless otherwise specified) and assign symbols to them.  The shortwave opacity,
\begin{equation}
\kappa_{\rm S} \equiv \frac{\int_{\rm S} \kappa_\nu J_\nu ~d\nu}{\int_{\rm S} J_\nu ~d\nu},
\label{eq:kappa_s}
\end{equation}
is assumed to be constant with the integration being taken over the range of shortwave frequencies.  Equation (\ref{eq:kappa_s}) defines the \emph{absorption mean} opacity (see \S3.2 of \citealt{mihalas}), which is defined to guarantee the correct total amount of (shortwave) energy absorption.  By contrast, the longwave opacity,
\begin{equation}
\kappa_{\rm L} \equiv \frac{\int_{\rm L} \kappa_\nu J_\nu ~d\nu}{\int_{\rm L} J_\nu ~d\nu} \approx \frac{\int_{\rm L} \kappa_\nu B_\nu ~d\nu}{\int_{\rm L} B_\nu ~d\nu},
\end{equation}
is closer to being a \emph{Planck mean} opacity defined to guarantee radiative equilibrium in the optically thin portions of the atmosphere (\S3.2 of \citealt{mihalas}).  The Planck/blackbody function is denoted by $B_\nu$.  The longwave opacity is in general a function of $m$, the cosine of the latitude $\mu = \cos\theta$ and the longitude $\phi$, i.e., $\kappa_{\rm L} = \kappa_{\rm L}(m,\mu,\phi)$.  For simplicity, we consider $\kappa_{\rm L} = \kappa_{\rm L}(m)$ only, while acknowledging the possibility that the strong temperature gradients inherent in hot Jupiter atmospheres may result in latitudinally and longitudinally varying longwave opacities.

Denoting the specific intensity by $I_\nu$ (with units of erg cm$^{-2}$ s$^{-1}$ Hz$^{-1}$ sr$^{-1}$), the radiative transfer equation is (e.g., pg. 35 of \citealt{mihalas})
\begin{equation}
\mu \frac{\partial I_\nu}{\partial m} = \frac{\kappa_\nu I_\nu}{\xi} - \kappa_\nu B_\nu - \frac{\kappa_\nu \left(1-\xi\right) J_\nu}{\xi}.
\label{eq:rt}
\end{equation}
Equation (\ref{eq:rt}) implicitly assumes either isotropic scattering or the collimated beam approximation (see Appendix \ref{append:two_stream}) where the forward and backward scattering probabilities are the same.  The moments of the radiative transfer equation are described by a pair of equations \citep{mihalas,hubeny03,guillot10},
\begin{equation}
\begin{split}
&\frac{\partial H_\nu}{\partial m} = \kappa_\nu \left( J_\nu - B_\nu \right), \\
&\frac{\partial K_\nu}{\partial m} = \frac{\kappa_\nu H_\nu}{\xi}.
\end{split}
\label{eq:main_equations}
\end{equation}
Since there are three unknowns ($J_\nu, H_\nu, K_\nu$) and only two equations, closure relations known as the Eddington approximations are required in order to obtain the solutions.  In the longwave, the first and second Eddington coefficients are respectively defined as 
\begin{equation}
{\cal E}_1 \equiv \frac{K_{\rm L}}{J_{\rm L}} = \frac{1}{3}, ~{\cal E}_2 \equiv \frac{H_{\rm L}}{J_{\rm L}} = \frac{1}{2},
\label{eq:eddington}
\end{equation}
and are assumed to be constant --- with their values chosen to be consistent with \cite{guillot10} --- to facilitate algebraic amenability.  We note that if the two-stream approximation is made --- which is typically the simplifying assumption adopted in three-dimensional simulations with radiative transfer (e.g., \citealt{showman09,hfp11}) --- then we instead have, e.g., ${\cal E}_2 = 1/\sqrt{3} \approx 0.58$ (see Appendix \ref{append:two_stream}).  In the shortwave, the first Eddington coefficient is equated to the square of the cosine of the latitude \citep{guillot10},
\begin{equation}
\mu^2 = \frac{K_{\rm S}}{J_{\rm S}}.
\label{eq:eddington2}
\end{equation}
We note that equation (\ref{eq:eddington2}) is a consequence of the collimated beam approximation which we make only in the shortwave (see Appendix \ref{append:two_stream}).

Let the heat transported by atmospheric circulation, as witnessed in three-dimensional simulations (e.g., \citealt{showman09,hmp11}), be $Q = Q(m,\mu,\phi)$, which has units of erg s$^{-1}$ g$^{-1}$ and is thus technically a specific luminosity of heat.  It can be related to the moments of the specific intensity by integrating the right-hand side of the first expression in equation (\ref{eq:main_equations}) over all frequencies,
\begin{equation}
\kappa_{\rm L} \left( J_{\rm L} - B \right) + \kappa_{\rm S} J_{\rm S} = Q,
\label{eq:q_equation}
\end{equation}
where we have $B = \int_{\rm L} B_\nu ~d\nu = \sigma_{\rm SB} T^4/\pi$, $\sigma_{\rm SB}$ denotes the Stefan-Boltzmann constant and we have assumed that the star and the hot Jupiter radiate negligibly in the longwave and shortwave, respectively.  Equation (\ref{eq:q_equation}) is equivalent to the energy equation used in general circulation models if one specializes to a static or time-independent situation.  By further integrating over column mass, it follows that
\begin{equation}
H = H_\infty - \tilde{Q}\left(m,\infty\right),
\label{eq:h}
\end{equation}
where $H_\infty$ is the value of $H$ evaluated when $m \rightarrow \infty$ and 
\begin{equation}
\tilde{Q}\left(m_1,m_2\right) \equiv \int^{m_2}_{m_1} Q\left(m^\prime, \mu, \phi \right) ~dm^\prime.
\end{equation}
Note that equation (\ref{eq:h}) is completely general: it requires no assumptions on the functional forms of either $\kappa_{\rm S}$ or $\kappa_{\rm L}$.  For convenience, the $\mu$- and $\phi$-dependences of $\tilde{Q}$ have been suppressed.

\subsection{Shortwave}

\begin{figure}
\begin{center}
\includegraphics[width=0.5\columnwidth]{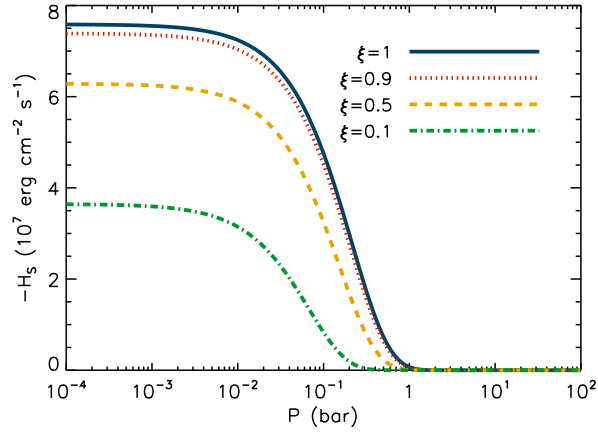}
\end{center}
\vspace{-0.2in}
\caption{Examples of the first moment of the shortwave intensity for $\mu=1$ and $T_{\rm irr}=2025$ K.}
\label{fig:hs}
\end{figure}

In the shortwave, equation (\ref{eq:main_equations}) reduces to
\begin{equation}
\begin{split}
&\frac{\partial H_{\rm S}}{\partial m} = \kappa_{\rm S} J_{\rm S}, \\
&\frac{\partial K_{\rm S}}{\partial m} = \frac{\kappa_{\rm S} H_{\rm S}}{\xi}.
\end{split}
\label{eq:shortwave}
\end{equation}
By using $\mu^2 = K_{\rm S}/J_{\rm S}$, one may rewrite the preceding pair of equations,
\begin{equation}
\begin{split}
&\frac{\partial^2J_{\rm S}}{\partial m^2} = \left( \frac{\kappa_{\rm S}}{\mu \sqrt{\xi}} \right)^2 J_{\rm S} + \frac{1}{\mu^2} \frac{\partial \left( \kappa_{\rm S}/\xi \right)}{\partial m} \left(\int^m_0 \kappa_{\rm S} J_{\rm S} ~dm^\prime + H_{\rm S_0} \right), \\
&\frac{\partial^2H_{\rm S}}{\partial m^2} = \left( \frac{\kappa_{\rm S}}{\mu \sqrt{\xi}} \right)^2 H_{\rm S} + \frac{1}{\mu^2} \frac{\partial \kappa_{\rm S}}{\partial m} \left( \int^m_0 \frac{\kappa_{\rm S} H_{\rm S}}{\xi} ~dm^\prime + \mu^2 J_{\rm S_0} \right). \\
\end{split}
\label{eq:shortwave2}
\end{equation}
Upon inspection of equation (\ref{eq:shortwave2}), it is evident that obtaining analytical solutions for $J_{\rm S}$ and $H_{\rm S}$ is a challenging task unless the assumptions of constant $\kappa_{\rm S}$ and $\xi$ are made.  With such assumptions and the boundary conditions that $J_{\rm S}=H_{\rm S}=0$ as $m \rightarrow \infty$, we obtain
\begin{equation}
\begin{split}
&J_{\rm S} = J_{\rm S_0} \exp{\left( -\frac{\kappa_{\rm S}m}{\mu \sqrt{\xi}} \right)} = J_{\rm S_0} \exp{\left( -\frac{\sqrt{\xi} \tau_{\rm S}}{\mu} \right)},\\
&H_{\rm S} = H_{\rm S_0} \exp{\left( -\frac{\kappa_{\rm S}m}{\mu \sqrt{\xi}} \right)} = H_{\rm S_0} \exp{\left( -\frac{\sqrt{\xi} \tau_{\rm S}}{\mu} \right)},
\end{split}
\label{eq:shortwave3}
\end{equation}
where $J_{\rm S_0}$ and $H_{\rm S_0}$ are the values of $J_{\rm S}$ and $H_{\rm S}$, respectively, evaluated at $m=0$ and whence
\begin{equation}
H_{\rm S} = - \mu \sqrt{\xi} J_{\rm S} \implies H_{\rm S_0} = - \mu \sqrt{\xi} J_{\rm S_0}.
\label{eq:h_s0}
\end{equation}
One may regard $\kappa_{\rm S}$ as a shortwave opacity which collectively describes the absorbing effects of the ``gas" and the ``haze" or ``cloud", since it is assumed to be constant.  Thus, the effects of absorption and scattering in the shortwave are described by the parameters $\kappa_{\rm S}$ and $\xi$, respectively.

Combining equation (\ref{eq:h_s0}) with the collimated beam approximation already employed in the formalism (see Appendix \ref{append:two_stream}), the Bond albedo may now be related to the scattering parameter (Figure \ref{fig:albedo_xi}),
\begin{equation}
{\cal A} = \frac{ 1 - \sqrt{\xi} }{ 1 + \sqrt{\xi} },
\label{eq:albedo_xi}
\end{equation}
which in turn allows us to make a physical connection between the strength of shortwave scattering and the temperatures deep in the atmosphere ($T_\infty$; equation [\ref{eq:t_infinity}]).  Note that we only consider forward and backward scattering in the collimated beam approximation (see Appendix \ref{append:two_stream}).

As an illustration, we show examples of $H_{\rm S}$ in Figure \ref{fig:hs}.  Two effects are apparent: the stellar photons are absorbed higher in the atmosphere as scattering becomes more important; and stronger scattering dilutes the incident stellar irradiation via the enhancement of the Bond albedo.  Both the shortwave photosphere (equation [\ref{eq:pressure_s}]) and photon deposition depth (equation [\ref{eq:deposition}]) shift to higher altitudes as scattering becomes a stronger effect (decreasing $\xi$), but the relative separation between them grows as $\propto 1/\sqrt{\xi}$.

\subsection{Longwave}

In the longwave, equation (\ref{eq:main_equations}) reduces to
\begin{equation}
\begin{split}
&\frac{\partial H_{\rm L}}{\partial m} = \kappa_{\rm L} \left( J_{\rm L} - B \right) = Q - \kappa_{\rm S} J_{\rm S}, \\
&\frac{\partial K_{\rm L}}{\partial m} = \kappa_{\rm L} H_{\rm L},
\end{split}
\label{eq:longwave}
\end{equation}
where the second equality in the first expression follows from the assumption of energy conservation as described in equation (\ref{eq:q_equation}).  Integrating over the column mass (per unit area) for the first expression in equation (\ref{eq:longwave}), we obtain
\begin{equation}
H_{\rm L} = H_{\rm L_0} + H_{\rm S_0} \left[ 1 - \exp{\left( -\frac{\kappa_{\rm S}m}{\mu \sqrt{\xi}} \right)} \right] + \tilde{Q}\left( 0, m \right).
\label{eq:hl1}
\end{equation}
In deriving the preceding expression, we have made use of equation (\ref{eq:h_s0}) which assumes a constant shortwave opacity.  However, equation (\ref{eq:hl1}) makes no assumption on the functional form of $\kappa_{\rm L}$.  The quantity $H_{\rm L_0}$ is the value of $H_{\rm L}$ evaluated at $m=0$. Using equation (\ref{eq:h}), it follows that
\begin{equation}
H_{\rm L} = H_\infty - H_{\rm S_0} \exp{\left( -\frac{\kappa_{\rm S}m}{\mu \sqrt{\xi}} \right)} - \tilde{Q}\left( m, \infty \right).
\label{eq:hl2}
\end{equation}

Combining equation (\ref{eq:eddington}), the second equation in (\ref{eq:longwave}) and equation (\ref{eq:hl2}) yields
\begin{equation}
J_{\rm L} = J_{\rm L_0} + \frac{1}{{\cal E}_1} \int^m_0 \kappa_{\rm L} \left[  H_\infty - H_{\rm S_0} \exp{\left( -\frac{\kappa_{\rm S}m^\prime}{\mu \sqrt{\xi}} \right)} - \tilde{Q}\left( m, \infty \right) \right] ~dm^\prime,
\label{eq:jl}
\end{equation}
where $J_{\rm L_0}$ is the value of $J_{\rm L}$ evaluated at $m=0$ and may be eliminated in favour of other quantities using equations (\ref{eq:eddington}) and (\ref{eq:hl2}),
\begin{equation}
J_{\rm L_0} = \frac{1}{{\cal E}_2} \left[ H_\infty - H_{\rm S_0} - \tilde{Q}\left( 0, \infty \right) \right].
\end{equation}

Our goal is to obtain an expression in terms of only $B$, $H_\infty$ and $H_{\rm S_0}$.  To this end, equation (\ref{eq:jl}) may be rewritten using equation (\ref{eq:q_equation}),
\begin{equation}
B = H_\infty \left( \frac{1}{{\cal E}_2} + \frac{1}{{\cal E}_1} \int^m_0 \kappa_{\rm L} ~dm^\prime \right) - H_{\rm S_0} \left[ \frac{1}{{\cal E}_2} + \frac{1}{{\cal E}_1} \int^m_0 \kappa_{\rm L} \exp{\left( -\frac{\kappa_{\rm S}m^\prime}{\mu \sqrt{\xi}} \right)} ~dm^\prime \right] + \frac{\kappa_{\rm S}}{\mu \sqrt{\xi} \kappa_{\rm L}} \exp{\left( -\frac{\kappa_{\rm S}m}{\mu \sqrt{\xi}} \right)} + {\cal Q},
\label{eq:b}
\end{equation}
where we have collected all of the terms involving $Q$ into the following quantity,
\begin{equation}
{\cal Q} \equiv - \frac{Q}{\kappa_{\rm L}} - \frac{1}{{\cal E}_1} \int^m_0 \kappa_{\rm L} ~\tilde{Q}\left( m^\prime, \infty \right) ~dm^\prime - \frac{\tilde{Q}\left( 0, \infty \right)}{{\cal E}_2}.
\label{eq:q_heat}
\end{equation}
We have defined ${\cal Q}<0$ because it is the heat flux --- as a function of depth, latitude and longitude, i.e., ${\cal Q} = {\cal Q}(m,\mu,\phi)$ --- transported by horizontal winds from the day to the night side of a hot Jovian atmosphere \citep{sg02}.  In the limit of a constant longwave opacity (i.e., $d\kappa_{\rm L}/dm=0$) and no shortwave scattering ($\xi=1$), equation (\ref{eq:b}) reduces to equation (41) of \cite{guillot10} when two typographical errors in the latter are corrected: one should obtain $-Q/\kappa_{\rm L}$ instead of $-Q$ and the terms involving $\tilde{Q}(m^\prime,\infty)$ are missing the ${\cal E}_1^{-1}$ coefficient.

\subsection{Global-mean temperature-pressure profile}

As realized by \cite{guillot10}, the pair of first moments in equation (\ref{eq:b}) have clear physical interpretations:
\begin{equation}
H_\infty = \frac{\sigma_{\rm SB} T^4_{\rm int}}{4 \pi}, ~H_{\rm S_0} = - \frac{\mu \sigma_{\rm SB} T^4_{\rm irr}}{4 \pi}.
\label{eq:h_meaning}
\end{equation}
The quantity $T_{\rm int}$ is the blackbody-equivalent temperature associated with the internal heat flux, while the irradiation temperature is
\begin{equation}
T_{\rm irr} = T_\star \left( \frac{R_\star}{a} \right)^{1/2} \left( 1 - {\cal A} \right)^{1/4} \approx 1900 \mbox{ K} ~\left( \frac{T_\star}{6000 \mbox { K}} \right) \left( \frac{R_\star/a}{0.1} \right)^{1/2} \left[ \frac{2\sqrt{\xi}}{\left(1+\sqrt{\xi}\right)} \right]^{1/4},
\end{equation}
where $T_\star$ is the stellar effective temperature, $R_\star$ is the stellar radius and $a$ is the spatial separation between the hot Jupiter and the star.  The albedo integrated over all shortwave frequencies is denoted by ${\cal A}$ and may be regarded as the Bond albedo (see \S3.4 of \citealt{seager10} for a discussion of albedos).  Note that the equilibrium temperature, in the absence of day-night heat redistribution, is $T_{\rm eq} = T_{\rm irr}/\sqrt{2}$.  The quantity $H_{\rm S_0}$ is negative because the incoming stellar irradiation travels downwards into the atmosphere.  For most values of the scattering parameter ($\xi \gtrsim 0.1$), the coefficient $(1-{\cal A})^{1/4}= [2\sqrt{\xi}/(1+\sqrt{\xi})]^{1/4}$ is typically close to being unity (Figure \ref{fig:albedo_xi}), which implies that the temperature at depth $T_\infty$ is insensitive to changes in the Bond albedo (cf. through $T_{\rm eq}$ in equation [\ref{eq:t_infinity}]) unless ${\cal A} \gtrsim 0.5$.

Substituting equation (\ref{eq:h_meaning}) into equation (\ref{eq:b}), we obtain
\begin{equation}
T^4 = \frac{T^4_{\rm int}}{4} \left( \frac{1}{{\cal E}_2} + \frac{1}{{\cal E}_1} \int^m_0 \kappa_{\rm L} ~dm^\prime  \right) + \frac{T^4_{\rm irr}}{4} \left[ \frac{\kappa_{\rm S}}{\sqrt{\xi} \kappa_{\rm L}}  \exp{\left( -\frac{\kappa_{\rm S}m}{\mu \sqrt{\xi}} \right)} + \frac{\mu}{{\cal E}_2} + \frac{\mu}{{\cal E}_1} \int^m_0 \kappa_{\rm L}  \exp{\left( -\frac{\kappa_{\rm S}m^\prime}{\mu \sqrt{\xi}} \right)} ~dm^\prime \right] + \frac{\pi {\cal Q}}{\sigma_{\rm SB}}.
\label{eq:t4}
\end{equation}
In the limit of a constant longwave opacity, the preceding expression reduces to equation (43) of \cite{guillot10} when a typographical in the latter is corrected for: instead of $\pi Q/\sigma_{\rm SB}$, one should have $-\pi Q/\sigma_{\rm SB} \kappa_{\rm L}$.  The reduction to equation (43) of \cite{guillot10} also requires the use of the identity (see Appendix \ref{append:identity}),
\begin{equation}
\frac{\partial}{\partial m^\prime} \tilde{Q}\left( m^\prime, \infty \right) = -Q\left(m^\prime, \mu, \phi \right),
\label{eq:q_heat2}
\end{equation}
towards evaluating the second term in equation (\ref{eq:q_heat}) via integration by parts.

For an arbitrary function ${\cal X}$, its global-mean counterpart is
\begin{equation}
\bar{{\cal X}} \equiv \frac{1}{2\pi} \int^{2\pi}_0 \int^1_0 {\cal X} ~d\mu ~d\phi,
\label{eq:averaging}
\end{equation}
with the assumption that there is latitudinal symmetry about the equator ($\theta=0^\circ$).  For the terms involving the internal heat flux ($\sigma_{\rm SB} T^4_{\rm int}$), we assume azimuthal symmetry, i.e., the day and night sides are the same.  For the terms involving the irradiation temperature ($T_{\rm irr}$), we are strictly speaking performing only a hemispherical averaging --- the integration over $\phi$ is only non-zero for $\pi$ radians over the day side, consistent with the theoretical expectation that hot Jupiters are tidally locked (at least in the case of circular orbits).  If one wishes to compute the temperature-pressure profile only on the day-side of a hot Jovian atmosphere, then the integration over $\phi$ needs only to be performed from 0 to $\pi$.\footnote{In other words, the term in equation (\ref{eq:t4_global}) involving $T_{\rm int}^4$ has a coefficient of 1/8 and not 1/4.}  Following \cite{guillot10}, we assert that
\begin{equation}
\frac{1}{2\pi} \int^{2\pi}_0 \int^1_0 {\cal Q} ~d\mu ~d\phi = 0.
\label{eq:no_heat}
\end{equation}
The three-dimensional simulations of atmospheric circulation of \cite{hfp11} demonstrate that the approximation in equation (\ref{eq:no_heat}) breaks down at $P \lesssim 10$ bar, where the simulated temperature-pressure profile (in dynamical-radiative equilibrium) is cooler by $\sim 50$ K, at least for their models of HD 209458b, due to the conversion of heat into mechanical energy via the creation of $\sim 1$ km s$^{-1}$ horizontal winds.  Since our goal is to derive temperature-pressure profiles in radiative equilibrium and \emph{not} full dynamical-radiative equilibrium, we retain this approximation.  One may regard our approach as being analogous to the ``no redistribution" models presented in \cite{hansen08}.  Thus, we get
\begin{equation}
\bar{T}^4 = \frac{T^4_{\rm int}}{4} \left( \frac{1}{{\cal E}_2} + \frac{1}{{\cal E}_1} \int^m_0 \kappa_{\rm L} ~dm^\prime  \right) + \frac{T^4_{\rm irr}}{8} \left[ \frac{1}{2{\cal E}_2} + \frac{\gamma}{\sqrt{\xi}} ~E_2\left(\frac{\kappa_{\rm S} m^\prime}{\sqrt{\xi}}\right) + \frac{1}{{\cal E}_1} \int^m_0 \kappa_{\rm L} ~E_3\left( \frac{\kappa_{\rm S} m^\prime}{\sqrt{\xi}} \right) ~dm^\prime \right],
\label{eq:t4_global}
\end{equation}
where we have explicitly stated the arguments of the exponential integrals and defined
\begin{equation}
\gamma \equiv \frac{\kappa_{\rm S}}{\kappa_{\rm L}}, ~\tau = \tau\left(m\right) \equiv \kappa_{\rm L} m.
\end{equation}
We call the quantity $\bar{T}$ the ``global-mean temperature", but strictly speaking it is the 1/4-root of the global-mean flux since the averaging is taken over $T^4$ and not $T$.  The exponential integral of the $j$-th order is described by the following expressions \citep{aw}:
\begin{equation}
\begin{split}
&E_j \left( x \right) = \int^\infty_1 y^{-j} \exp{\left(-xy\right)} ~dy,\\
&E_{j+1} \left( x \right) = \frac{1}{j} \left[ \exp{\left(-x \right)} - x E_j\left(x\right) \right].
\end{split}
\end{equation}
If we set $\kappa_{\rm L}$ to be constant, use $\xi=1$ and adopt the numerical values of the Eddington coefficients as stated in equation (\ref{eq:eddington}), then equation (\ref{eq:t4_global}) reduces to equation (49) of \cite{guillot10}.  Note that none of the typographical errors previously described affect equation (49) of \cite{guillot10}, because of the assertion in equation (\ref{eq:no_heat}).

In summary, equation (\ref{eq:t4_global}) is the general expression for the global-mean temperature, as a function of depth, in a hot Jovian atmosphere, assuming constant shortwave opacity but allowing for an arbitrary functional form for the longwave opacity.  In the following section, we will examine various forms of this equation with $\kappa_{\rm L} = \kappa_{\rm L}(m)$ specified.

\section{Applications of the formalism}
\label{sect:applications}

To proceed further, we first need to relate the column mass (per unit area) $m$ to the pressure $P$.  If we specify $z$ to be the vertical coordinate as measured from the top of the atmosphere (at $z=0$) downwards, then hydrostatic balance demands
\begin{equation}
\frac{dP}{dz} = \rho g,
\end{equation}
where $\rho$ is the mass density of the atmospheric fluid and $g \sim 10^3$ cm s$^{-1}$.  By assuming that $g$ remains constant over the extent of the (thin) atmosphere and defining $dm \equiv \rho dz$, we obtain
\begin{equation}
P = mg.
\end{equation}
For the purpose of comparison to three-dimensional simulations, which require the specification of a bottom (with a pressure $P_0$) for the computational domain, we define the column mass per unit area at the bottom to be
\begin{equation}
m_0 \equiv P_0/g.
\end{equation}
Therefore, we have $P/P_0 = m/m_0$.  We adopt $P_0=100$ bar throughout this study.

Upon specifying a functional form for $\kappa_{\rm L} = \kappa_{\rm L}(m)$, the simplification of equation (\ref{eq:t4_global}) only involves the evaluation of the double integral,
\begin{equation}
{\cal I} = \frac{1}{{\cal E}_1} \int^\infty_1 x^{-3} \int^m_0 \kappa_{\rm L}\left(m^\prime\right) ~\exp{\left( - \frac{\kappa_{\rm S} m^\prime x}{\sqrt{\xi}} \right)} ~dm^\prime ~dx.
\end{equation}
For example, when $\kappa_{\rm L}$ is constant, we have
\begin{equation}
{\cal I} = \frac{\sqrt{\xi}}{3\gamma {\cal E}_1} \left[ 1 - 3 E_4 \left( \frac{\gamma \tau}{\sqrt{\xi}} \right) \right].
\end{equation}

\subsection{Collision-induced absorption}

\begin{figure}
\begin{center}
\includegraphics[width=0.5\columnwidth]{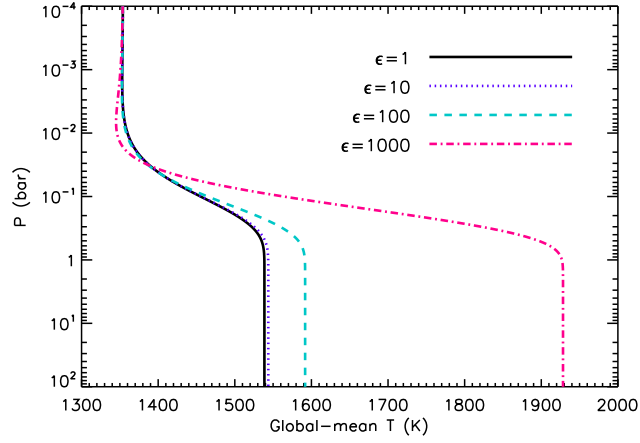}
\end{center}
\vspace{-0.2in}
\caption{Temperature-pressure profiles with $\xi=1$ and various values of $\epsilon$, which parametrizes the effect of collision-induced absorption in the longwave.}
\label{fig:cia}
\end{figure}

As previously noted in \cite{guillot10} and \cite{hfp11}, the assumption of a constant longwave opacity breaks down at high pressures because of the effect of collision-induced absorption \citep{herzberg52,pierrehumbert}.\footnote{Note that collision-induced absorption and pressure broadening are different effects: the former produces continuum absorption, while the latter is associated with the broadening of absorption lines.}  We then have $\kappa_{\rm L} \propto P \propto m$.  Following \cite{fhz06} and \cite{hfp11}, we adopt the following functional form for the longwave optical depth,
\begin{equation}
\tau_{\rm L}\left(P\right) = \tau_0 \left[ \frac{P}{P_0} + \left( \epsilon - 1 \right) \left(\frac{P}{P_0}\right)^2 \right],
\label{eq:tau_L}
\end{equation}
such that $\tau_{\rm L} \approx \tau_0 P/P_0$ and $\tau_{\rm L} = \epsilon \tau_0$ at the top and bottom of the model atmosphere, respectively.  The dimensionless quantity $\epsilon$ is the correction factor to the longwave optical depth due to the effect of collision-induced absorption, which becomes important when $P \gtrsim P_0/(\epsilon-1)$.  The functional form for the longwave opacity is
\begin{equation}
\kappa_{\rm L}\left(m\right) = \kappa_0 \left[ 1 + 2\left( \epsilon - 1 \right) \left( \frac{m}{m_0} \right) \right] \approx
\begin{cases}
\kappa_0, & m \ll m_0 ,\\
\kappa_0 \left( 2\epsilon - 1 \right), & m = m_0,
\end{cases}
\label{eq:kappa_L}
\end{equation}
which guarantees that equation (\ref{eq:tau_L}) is recovered when we evaluate
\begin{equation}
\tau_{\rm L}\left(m\right) = \int^m_0 \kappa_{\rm L}\left(m^\prime\right) ~dm^\prime
\end{equation}
and use $m/m_0 = P/P_0$.  The normalization for the longwave optical depth in the absence of collision-induced absorption is $\tau_0 \equiv \kappa_0 m_0$.
It follows that equation (\ref{eq:t4_global}) reduces to
\begin{equation}
\begin{split}
\bar{T}^4 =& \frac{3 T^4_{\rm int}}{4} \left( \frac{2}{3} + \tau_{\rm L}  \right) + \frac{3 T^4_{\rm eq}}{4} \left\{ \frac{2}{3} + \frac{2 \sqrt{\xi}}{3\gamma_0} \left[ 1 + \exp{\left(-\frac{\gamma\tau}{\sqrt{\xi}} \right)} \left( \frac{\gamma\tau}{2\sqrt{\xi}} - 1\right) \right] +  \frac{2\gamma}{3\sqrt{\xi}} ~E_2\left( \frac{\gamma \tau}{\sqrt{\xi}} \right) ~\left[ 1 - \frac{\tau^2}{2}\left( \frac{\gamma}{\gamma_0} \right) \right] \right\} \\
&+ \frac{3 T^4_{\rm eq} \xi}{4} \left( \frac{\epsilon-1}{\tau_0 \gamma^2_0} \right) \left[ 1 - \exp{\left(-\frac{\gamma\tau}{\sqrt{\xi}} \right)} - \frac{3 \gamma \tau}{\sqrt{\xi}} ~E_4\left( \frac{\gamma \tau}{\sqrt{\xi}} \right) \right] ,\\
\end{split}
\label{eq:t4_global_cia}
\end{equation}
where the functional dependences of the exponential integrals on $\gamma\tau/\sqrt{\xi}$ have been explicitly stated and
\begin{equation}
\gamma_0 \equiv \frac{\kappa_{\rm S}}{\kappa_0}.
\end{equation}
In the absence of clouds/hazes, atmospheres with and without temperature inversions are characterized by $\gamma_0 > 1$ and $\gamma_0 < 1$, respectively.  When $\epsilon=1$ and $\xi=1$, we obtain $\tau_{\rm L} = \tau$, $\gamma = \gamma_0$ and equation (\ref{eq:t4_global_cia}) reduces to equation (49) of \cite{guillot10}.  The temperature at depth ($T_{\rm int}=0$ K, $\tau \gg 1$) is
\begin{equation}
T_\infty \approx T_{\rm eq} \left[ \frac{1}{2} + \frac{\sqrt{\xi}}{2\gamma_0} + \frac{3 \xi \left(\epsilon-1\right)}{4 \gamma_0 \tau_{\rm S_0}} \right]^{1/4}.
\label{eq:t_infinity}
\end{equation}
The quantity $\tau_{\rm S_0}$ is the value of the shortwave optical depth, $\tau_{\rm S} = \kappa_{\rm S} m$, evaluated at $m=m_0$.  Figure \ref{fig:cia} shows examples of temperature-pressure profiles, with $\xi=1$, adopting the numbers used in \cite{hfp11} for their models of HD 209458b: $T_{\rm int}=0$ K, $T_{\rm eq}=1432$ K, $\kappa_{\rm S} = 0.006$ cm$^2$ g$^{-1}$, $\kappa_0 = 0.01$ cm$^2$ g$^{-1}$ ($\gamma_0=0.6$) and $\tau_{\rm S_0}=1401$.  \textit{It is apparent that larger values of $\epsilon$ correspond to higher temperatures at depth,} an effect which is similar to line blanketing in stellar atmospheres.  For $\epsilon=1000$, the temperatures at $P \sim 0.01$ bar appear cooler than the other cases because the enhanced opacity contribution from collision-induced absorption becomes noticeable at these pressures, similar to the cooling effect due to the presence of a cloud/haze layer (see \S\ref{subsect:constant}).  Elucidating the chemistry/physics which determines the exact value of $\epsilon$ is beyond the scope of the present study (see \S4.4.8 of \citealt{pierrehumbert} for a review of collision-induced absorption).  Rather, we demonstrate that the effect of collision-induced absorption can be approximately described using a a single parameter.  Note that the value of $T_\infty$ obtained, for an assumed value of $\epsilon$, depends on the value of $P_0$ adopted --- in other words, one needs to know the enhancement of the longwave optical depth, due to collision-induced absorption, at a reference pressure level (e.g., $P_0$).  Another way of interpreting the results in Figure {\ref{fig:cia} is that if $\epsilon \lesssim 100$ at $P_0 \sim 100$ bar, then collision-induced absorption is a minor effect (i.e., temperature differences $< 100$ K).  As further examples, we re-compute the values of $T_\infty$, for $\xi=1$, used in the simulations of \cite{hfp11}: for $P_0=220$ bar and setting $\epsilon = 10, 100, 1000$ and 2000, we have $T_\infty \approx 1541, 1564, 1749$ and 1903 K, respectively.

Finally, we note that the (simpler) generalization of Section 2 of \cite{guillot10} to consider equations (\ref{eq:tau_L}) and (\ref{eq:kappa_L}) was previously presented in \S4.3 of \cite{hfp11}.  Also, the approximations of constant $\kappa_{\rm S}$ and $\kappa_{\rm L} \propto P$ were previously adopted by \cite{ls10} for simulating the atmospheres of the gas giants in our Solar System.  Furthermore, \cite{hfp11} used the functional form in equation (\ref{eq:tau_L}) to study the atmospheres of hot Jupiters and were able to produce zonal-mean flow quantities (e.g., wind, temperature) which are similar to those published by \cite{showman09}, who used full opacity tables in their multi-wavelength calculations.

\subsection{Uniform cloud/haze layer with absorption and scattering}

\begin{figure}
\begin{center}
\includegraphics[width=0.5\columnwidth]{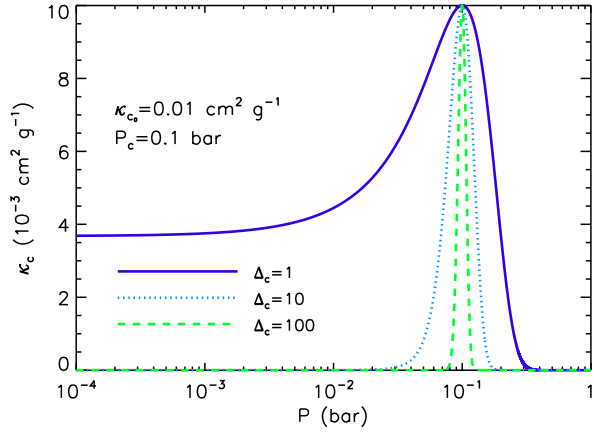}
\end{center}
\vspace{-0.2in}
\caption{Hypothetical cloud/haze decks parametrized using equation (\ref{eq:cloud_deck}).  Shown are three examples with different values of the deck thickness parameter.}
\label{fig:cloud_deck}
\end{figure}

In the simple case of a uniform cloud/haze layer, one can include the effects of both shortwave scattering ($\xi < 1$) and an extra, longwave opacity contribution $\kappa_{\rm c_0}$.  Equation (\ref{eq:t4_global}) then becomes
\begin{equation}
\begin{split}
\bar{T}^4 =& \frac{3 T^4_{\rm int}}{4} \left( \frac{2}{3} + \tau_{\rm L}  \right) + \frac{3 T^4_{\rm eq}}{4} \left\{ \frac{2}{3} + \frac{2 \sqrt{\xi}}{3\gamma_0} \left[ 1 + \exp{\left(-\frac{\gamma\tau}{\sqrt{\xi}} \right)} \left( \frac{\gamma\tau}{2\sqrt{\xi}} - 1\right) \right] +  \frac{2\gamma}{3\sqrt{\xi}} ~E_2\left( \frac{\gamma \tau}{\sqrt{\xi}} \right) ~\left[ 1 - \frac{\tau^2}{2}\left( \frac{\gamma}{\gamma_0} \right) \right] \right\} \\
&+ \frac{3 T^4_{\rm eq} \xi}{4} \left( \frac{\epsilon-1}{\tau_0 \gamma^2_0} \right) \left[ 1 - \exp{\left(-\frac{\gamma\tau}{\sqrt{\xi}} \right)} - \frac{3 \gamma \tau}{\sqrt{\xi}} ~E_4\left( \frac{\gamma \tau}{\sqrt{\xi}} \right) \right] + \frac{T^4_{\rm eq} \sqrt{\xi}}{2 \gamma_{\rm c}} \left[ 1 - 3 E_4\left( \frac{\gamma \tau}{\sqrt{\xi}} \right) \right],\\
\end{split}
\label{eq:t4_global_cloud0}
\end{equation}
where we have $\gamma_{\rm c} \equiv \kappa_{\rm S}/\kappa_{\rm c_0}$.  Equation (\ref{eq:t4_global_cloud0}) is essentially an analytical expression short of having to evaluate the exponential integrals.

In the longwave, we have made a distinction between the extra cloud/haze opacity $\kappa_{\rm c_0}$ and the opacity $\kappa_{\rm L}$ describing the rest of the atmosphere (the ``gas") because we assert the former to be constant while the latter is subjected to collision-induced absorption.  One may also choose to set $\kappa_{\rm c_0}=0$ cm$^2$ g$^{-1}$ and simply assimilate the longwave absorption properties of the haze/cloud into $\kappa_{\rm L}$.  In the shortwave, the opacity contribution due to the cloud/haze can likewise be added to that of the gas because the shortwave opacities are assumed to be constant.

\subsection{Purely absorbing cloud or haze deck}

The formation and dispersal of clouds or hazes is a fundamental, unsolved problem in the study of terrestrial climate science \citep{pierrehumbert}, brown dwarfs (e.g., \citealt{bht11}) and exoplanetary atmospheres.  To keep our formalism semi-analytical, we allow for the presence of a non-uniform cloud deck via the (ad hoc) function,
\begin{equation}
\kappa_{\rm c} = \kappa_{\rm c_0} \exp{\left[-\Delta_{\rm c} \left( 1 - \frac{P}{P_{\rm c}}  \right)^2 \right]}.
\label{eq:cloud_deck}
\end{equation}
Thus, our cloud deck is described by three free parameters: the opacity normalization $\kappa_{\rm c_0}$, the deck thickness parameter $\Delta_{\rm c}$ and the location of the deck $P_{\rm c}$.  Figure \ref{fig:cloud_deck} shows three examples of hypothetical cloud decks located at $P_{\rm c} = 0.1$ bar and with different deck thicknesses --- it is apparent that the larger the value of $\Delta_{\rm c}$, the thinner the cloud/haze deck.  In the limiting case of $\Delta_{\rm c}=0$, equation (\ref{eq:cloud_deck}) describes an extra opacity contribution which is constant throughout the atmosphere.  In the interest of algebraic amenability, we consider the cloud/haze deck to contribute an extra source of opacity only in the longwave; we set $\xi=1$ since a constant scattering opacity is inconsistent with an extra, non-uniform absorption opacity.  While being a simple starting point for non-uniform decks, real clouds or hazes probably do not exhibit this type of simple behaviour.  (See \citealt{pierrehumbert} for a review of the physics of clouds and their effects on the atmosphere.)  For example, clouds in brown dwarfs have been observationally inferred to be patchy \citep{artigau09}.  Nevertheless, equation (\ref{eq:cloud_deck}) provides us with a lucid vocabulary for describing cloud/haze decks: dense (higher $\kappa_{\rm c_0}$) versus tenuous (lower $\kappa_{\rm c_0}$), thick (higher $\Delta_{\rm c}$) versus thin (lower $\Delta_{\rm c}$), high (lower $P_{\rm c}$) versus low (higher $P_{\rm c}$).  The main motivation in this sub-section is to isolate the effect of an extra, non-uniform, infrared/longwave source of opacity.

The additional longwave optical depth due to the cloud/haze deck is
\begin{equation}
\tau_{\rm c} = \int^m_0 \kappa_{\rm c} ~dm^\prime = \frac{\kappa_{\rm c_0} m_{\rm c}}{2} \sqrt{\frac{\pi}{\Delta_{\rm c}}} \left[ \tilde{E}\left( \Delta^{1/2}_{\rm c} \right) - \tilde{E}\left( \Delta^{1/2}_{\rm c}\left(1-\frac{m}{m_{\rm c}} \right) \right) \right],
\label{eq:tau_c}
\end{equation}
where we have defined $m_c \equiv P_c/g$ and the error function is \citep{aw},
\begin{equation}
\tilde{E}\left(x\right) \equiv \frac{2}{\sqrt{\pi}} \int^x_0 \exp{\left(-y^2\right)} ~dy.
\end{equation}
Generalizing equation (\ref{eq:tau_L}), the longwave optical depth becomes
\begin{equation}
\tau_{\rm L} = \tau_0 \left[ \frac{P}{P_0} + \left( \epsilon - 1 \right) \left(\frac{P}{P_0}\right)^2 \right] + \tau_{\rm c}.
\label{eq:tau_L2}
\end{equation}

It follows that the global-mean temperature-pressure profile becomes
\begin{equation}
\begin{split}
\bar{T}^4 =& \frac{3 T^4_{\rm int}}{4} \left( \frac{2}{3} + \tau_{\rm L}  \right) + \frac{3 T^4_{\rm eq}}{4} \left\{ \frac{2}{3} + \frac{2}{3\gamma_0} \left[ 1 + \exp{\left(-\gamma\tau\right)} \left( \frac{\gamma\tau}{2} - 1\right) \right] +  \frac{2\gamma}{3} E_2\left(\gamma \tau\right) \left[ 1 - \frac{\tau^2}{2}\left( \frac{\gamma}{\gamma_0} \right) \right] + 2{\cal J} \right\} \\
&+ \frac{3 T^4_{\rm eq}}{4} \left( \frac{\epsilon-1}{\tau_0 \gamma^2_0} \right) \left[ 1 - \exp{\left(-\gamma\tau\right)} - 3 \gamma \tau E_4\left(\gamma \tau\right) \right],\\
\end{split}
\label{eq:t4_global_cloud}
\end{equation}
with the integral ${\cal J}$ being described by
\begin{equation}
\begin{split}
&{\cal J}\left(m\right) \equiv \int^\infty_1 x^{-3} ~{\cal J}_0\left(x,m\right) ~dx, \\
&{\cal J}_0\left(x,m\right) \equiv \frac{\kappa_{\rm c_0} m_{\rm c}}{2} \sqrt{\frac{\pi}{\Delta_{\rm c}}} ~\exp{\left[ \frac{\left( \kappa_{\rm S} m_{\rm c} \right)^2}{4 \Delta_{\rm c}} x^2 - \kappa_{\rm S} m_{\rm c} x \right] ~\left[ \tilde{E}\left( \Delta^{1/2}_{\rm c} - \frac{\kappa_{\rm S} m_{\rm c} x}{2 \Delta_{\rm c}^{1/2}} \right) - \tilde{E}\left( \Delta^{1/2}_{\rm c}\left( 1 - \frac{m}{m_{\rm c}}\right) - \frac{\kappa_{\rm S} m_{\rm c} x}{2 \Delta_{\rm c}^{1/2}}\right) \right]},
\end{split}
\end{equation}
where the arguments of the exponential integrals and error functions have again been explicitly written out.  The integral ${\cal J}$ does not have a general analytical solution, needs to be evaluated numerically and accounts for the warming effect of the clouds in the lower atmosphere.

\subsection{Longwave opacity with power-law term}

For completeness, we note that if the longwave optical depth and opacity have functional forms consisting of linear and power-law terms,
\begin{equation}
\begin{split}
&\tau_{\rm L}\left(P\right) = \tau_0 \left[ \frac{P}{P_0} + \left( \epsilon - 1 \right) \left(\frac{P}{P_0}\right)^{n+1} \right],\\
&\kappa_{\rm L}\left(m\right) = \kappa_0 \left[ 1 + \left(n+1\right)\left( \epsilon - 1 \right) \left( \frac{m}{m_0} \right)^n \right],
\end{split}
\end{equation}
where $n>0$ is dimensionless, then the simplification of equation (\ref{eq:t4_global}) requires the numerical evaluation of the integral,
\begin{equation}
{\cal I} = \frac{\sqrt{\xi}}{3\gamma_0 {\cal E}_1} \left[ 1 - 3 E_4 \left( \frac{\gamma \tau}{\sqrt{\xi}} \right) \right] + \frac{\kappa_0 \left( n + 1 \right) \left( \epsilon - 1 \right)}{{\cal E}_1} \int^m_0 \left( \frac{m^\prime}{m_0} \right)^n E_3\left( \frac{\kappa_{\rm S} m^\prime}{\sqrt{\xi}} \right) ~dm^\prime.
\end{equation}

\section{Specific examples of temperature-pressure profiles}
\label{sect:examples}

All of the results in this section assume $T_{\rm int}=0$ K.

\subsection{Absorption only}

All of the results in this sub-section assume $\xi=1$.

\subsubsection{Profiles with constant cloud/haze opacity}
\label{subsect:constant}

Using equation (\ref{eq:t4_global_cloud0}), Figure \ref{fig:cloud1} shows examples of temperature-pressure profiles for $\kappa_{\rm c} = \kappa_{\rm c_0}$ and adopting parameters similar to those for the hot Jupiter HD 209458b: $T_{\rm eq}=1432$ K, $g=9.42$ m s$^{-2}$, $\kappa_0 = 0.01$ cm$^2$ g$^{-1}$.  For illustration, we have set ${\cal A}=0$ and $\epsilon=100$ --- the former follows immediately from $\xi=1$ and equation (\ref{eq:albedo_xi}), while the latter implies that collision-induced absorption is the dominant source of the longwave optical depth at $P \gtrsim 2$ bar.  Higher values of $\epsilon$ simply translate into higher values of the temperatures at depth $T_\infty$ (Figure \ref{fig:cia}).

The left and right panels of Figure \ref{fig:cloud1} show the cases of $\gamma_0=0.6$ and 20, respectively.  \textit{The effect of the cloud/haze, present throughout the atmosphere, is to cool and warm the upper and lower atmospheres, respectively.}  As the cloud opacity ($\kappa_{\rm c_0}$) is increased, the longwave photosphere located at $P_{\rm L} \sim g/(\kappa_0+\kappa_{\rm c_0})$ shifts to higher altitudes.  For $\gamma_0=0.6$ (left panel), the lower atmosphere ($P \gtrsim 0.1$ bar) has a temperature which is generally higher than the equilibrium temperature of $T_{\rm eq}=1432$ K --- the analogue of the greenhouse effect for hot Jupiters.  For $\gamma_0=20$, the effect of the cloud/haze is still to cool and warm the upper and lower atmospheres, respectively, but the lower atmosphere now has a temperature which is generally \emph{lower} than the equilibrium temperature --- the analogue of the anti-greenhouse effect for hot Jupiters.  (See \S4.3.5 of \citealt{pierrehumbert} for a discussion of the greenhouse and anti-greenhouse effects.)

\begin{figure}
\begin{center}
\includegraphics[width=0.48\columnwidth]{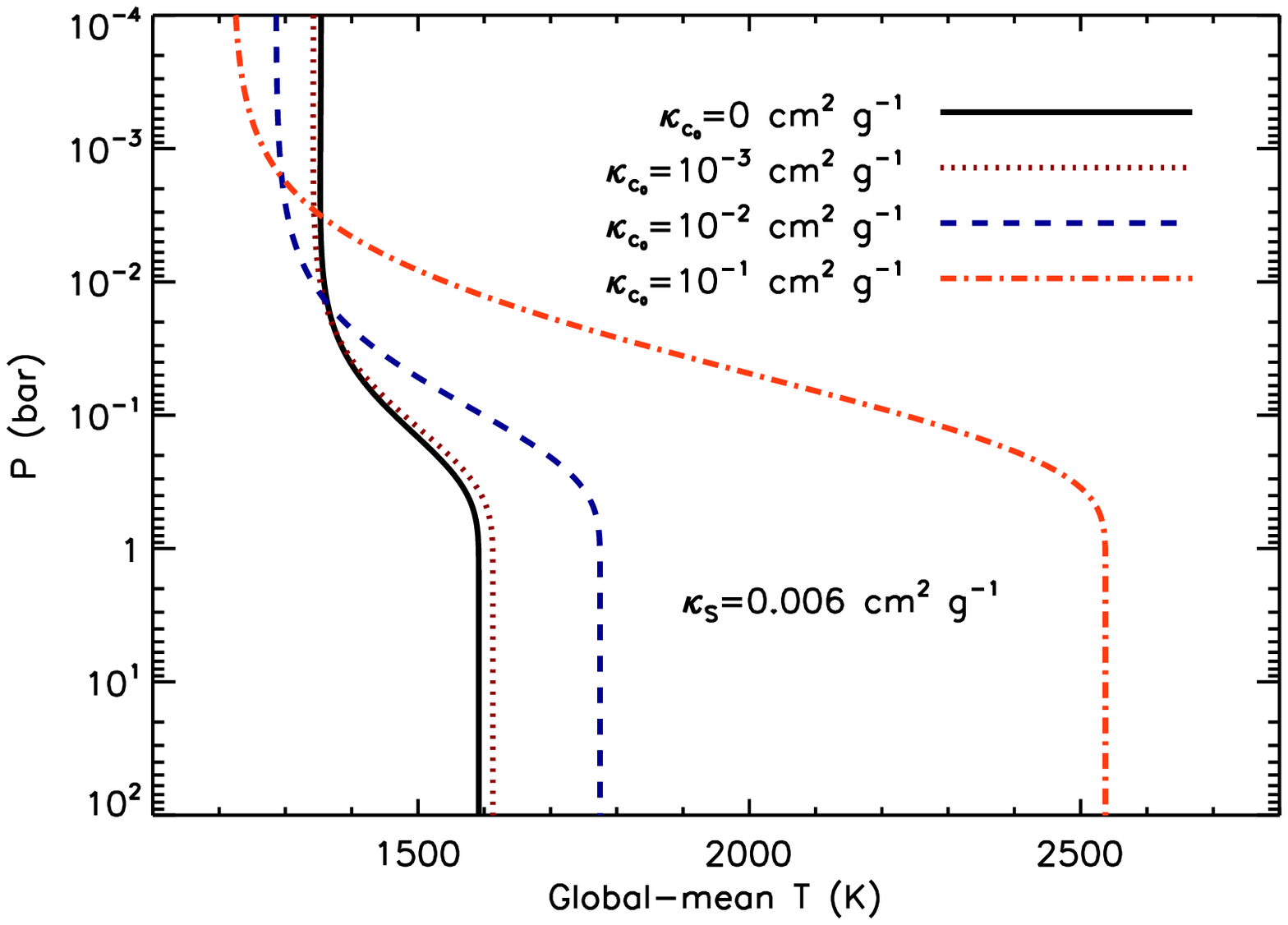}
\includegraphics[width=0.48\columnwidth]{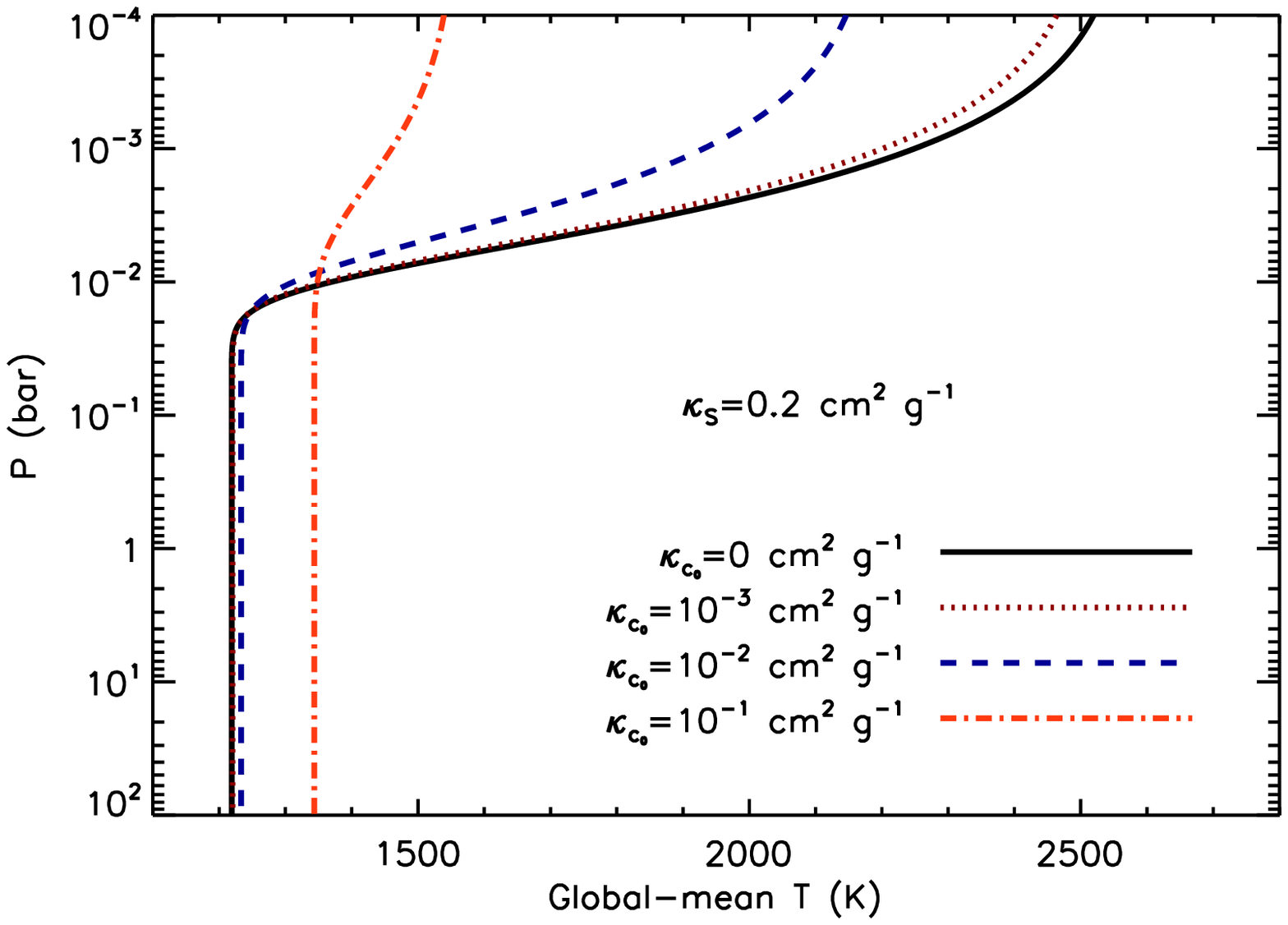}
\end{center}
\vspace{-0.2in}
\caption{Global-mean temperature-pressure profiles in the limiting case of constant cloud opacity ($\kappa_{\rm c} = \kappa_{\rm c_0}$) and $\xi=1$.  Left panel: $\kappa_{\rm S} = 0.006$ cm$^2$ g$^{-1}$ ($\gamma_0=0.6$).  Right panel: $\kappa_{\rm S} = 0.2$ cm$^2$ g$^{-1}$ ($\gamma_0=20$).  The equilibrium temperature of the hot Jupiter considered here is $T_{\rm eq}=1432$ K.}
\label{fig:cloud1}
\end{figure}

Previously, it was pointed out by \cite{hubeny03}, \cite{hansen08} and \cite{guillot10} that temperature inversions occur in an atmosphere when $\gamma_0>1$.  When a uniform cloud layer is present, the condition for a temperature inversion being present becomes
\begin{equation}
\gamma_0 > \mbox{max}\left\{ 0, \left( 1- \gamma^{-1}_{\rm c} \right)^{-1} \right\}.
\label{eq:inversion_condition}
\end{equation}
In other words, if $\gamma_{\rm c}>1$, then whether $\gamma_0 > 1$ becomes irrelevant.  When clouds are absent, we have $\gamma^{-1}_{\rm c} = 0$.

\subsubsection{Profiles with Gaussian cloud/haze decks}
\label{subsect:gaussian}

\begin{figure}
\begin{center}
\includegraphics[width=0.48\columnwidth]{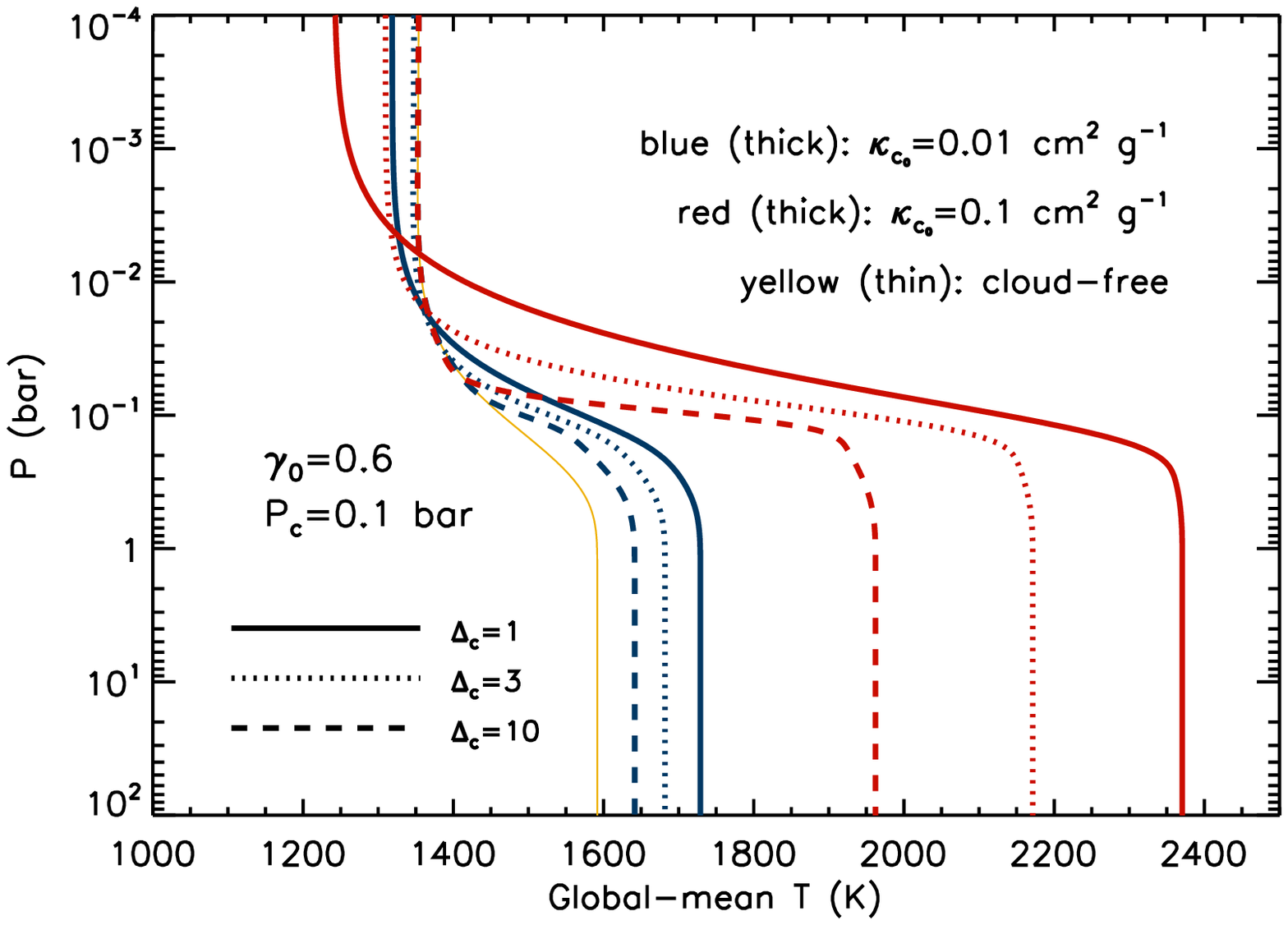}
\includegraphics[width=0.48\columnwidth]{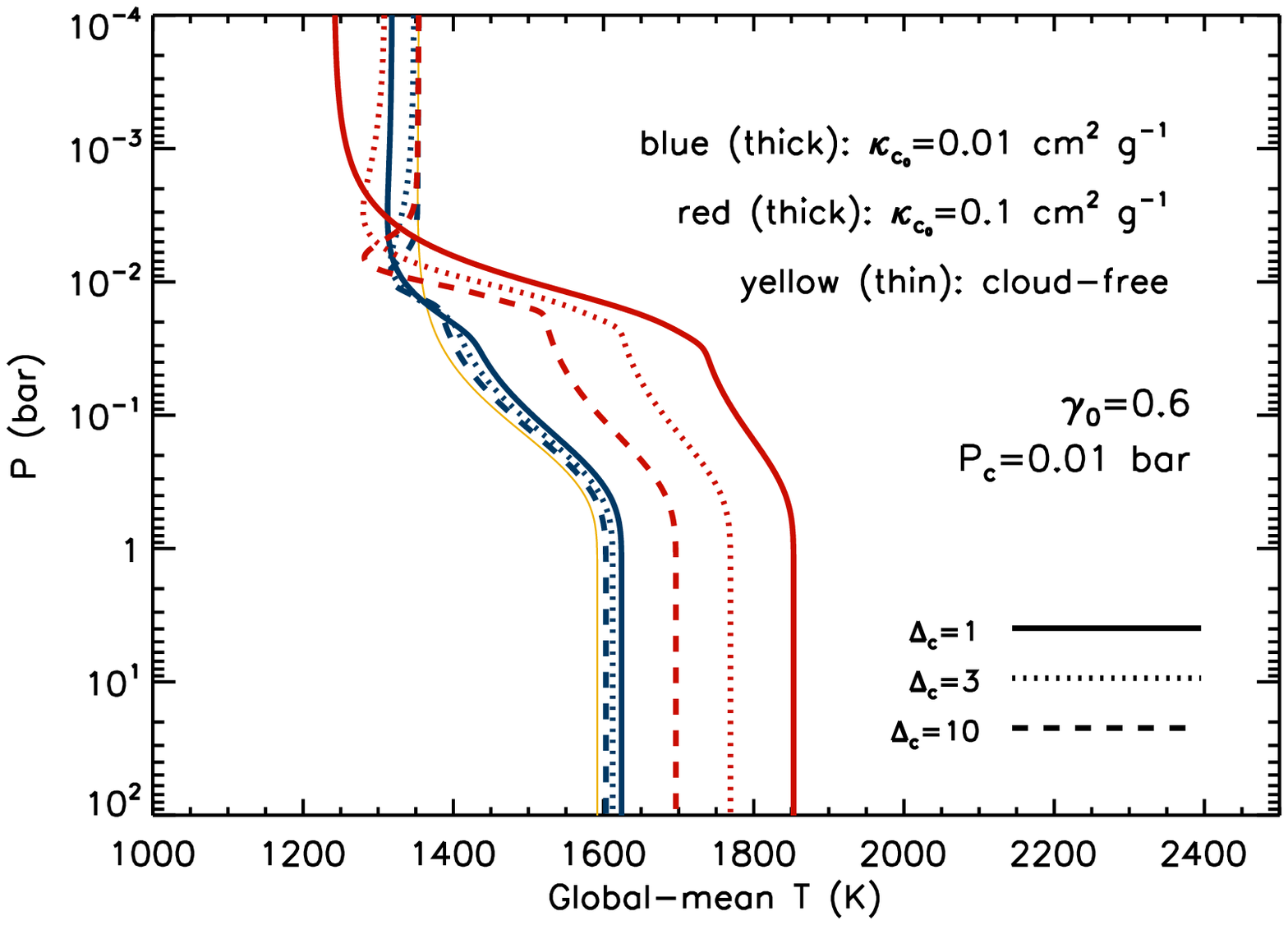}
\includegraphics[width=0.48\columnwidth]{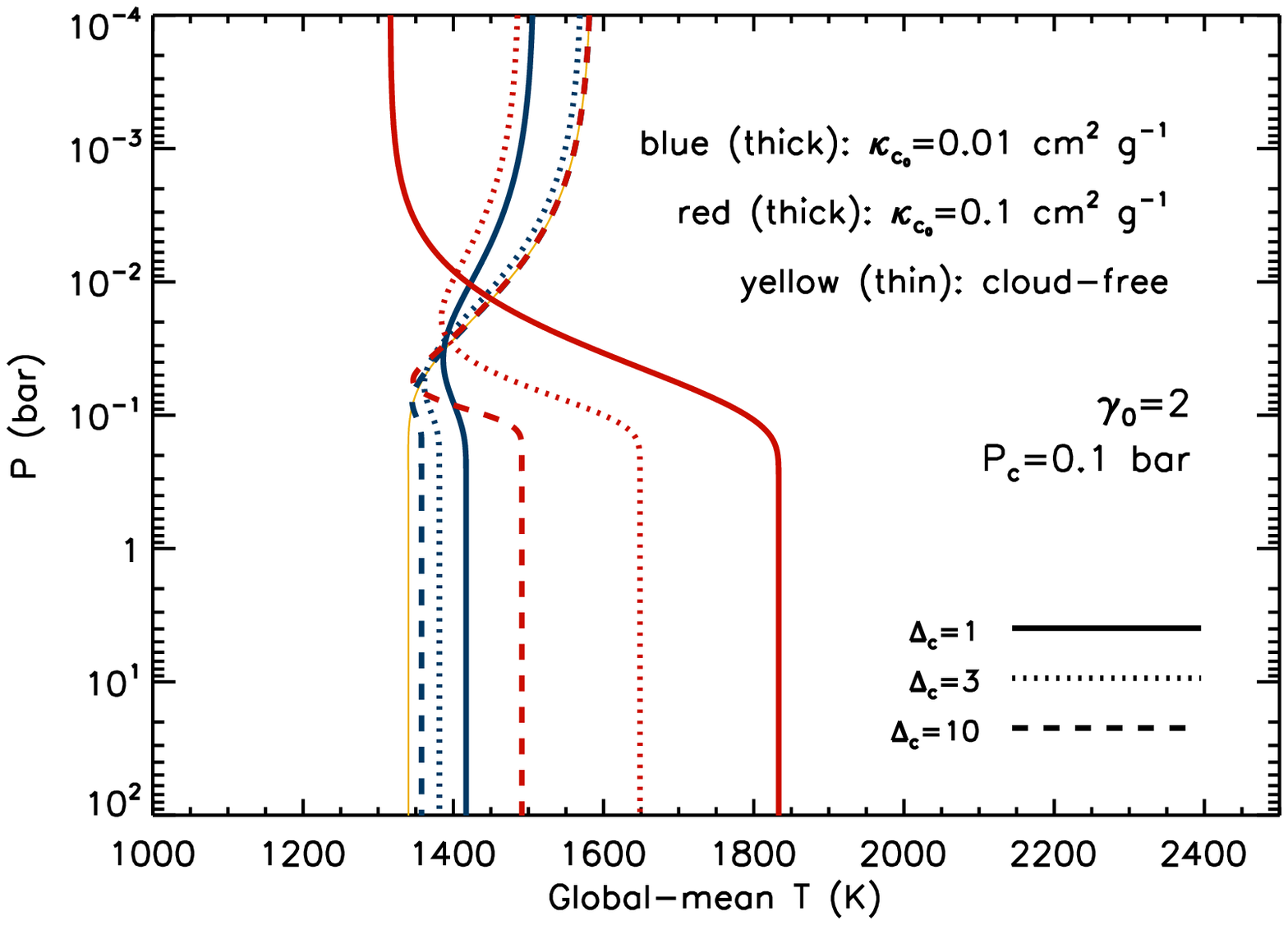}
\includegraphics[width=0.48\columnwidth]{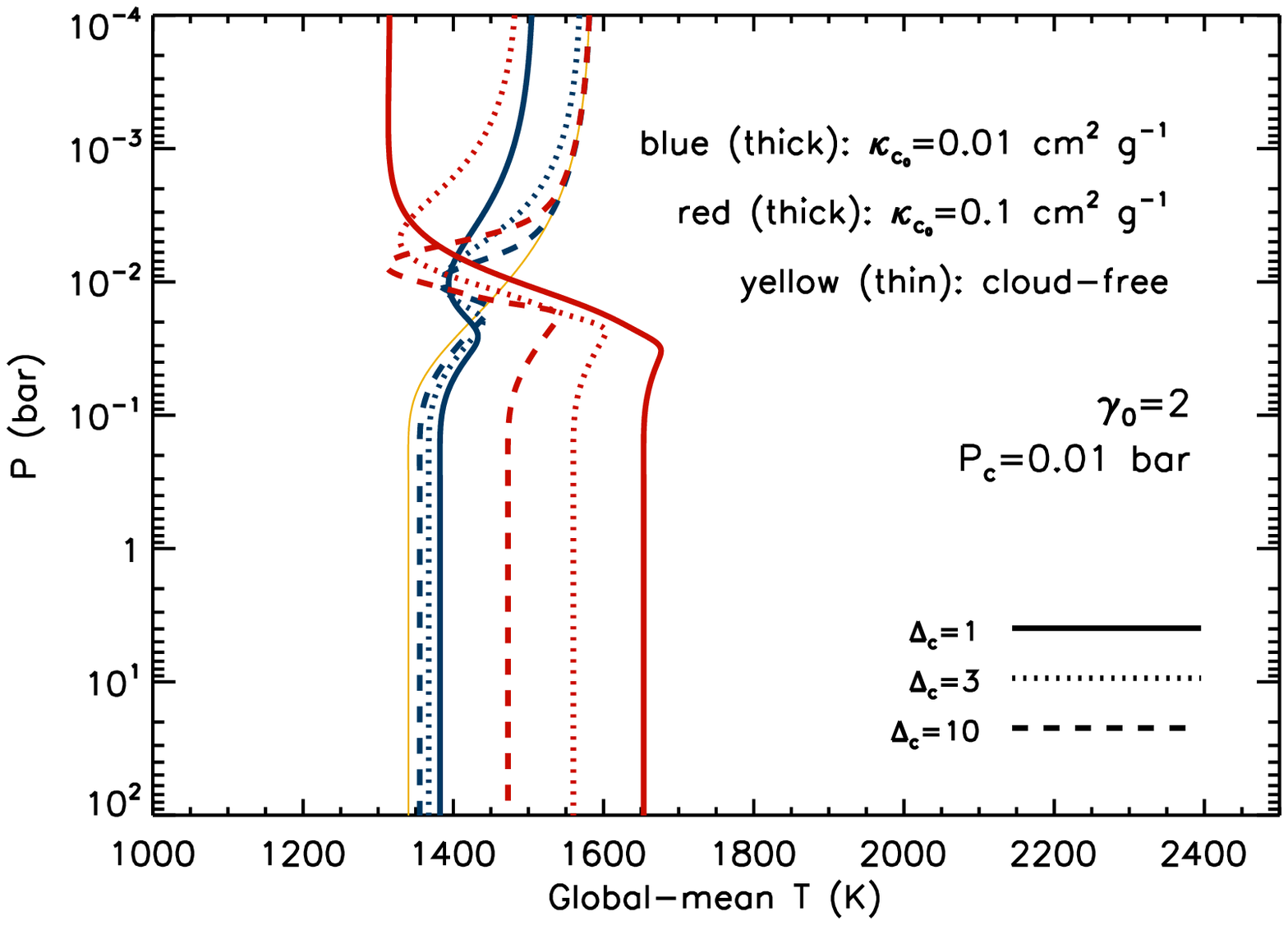}
\includegraphics[width=0.48\columnwidth]{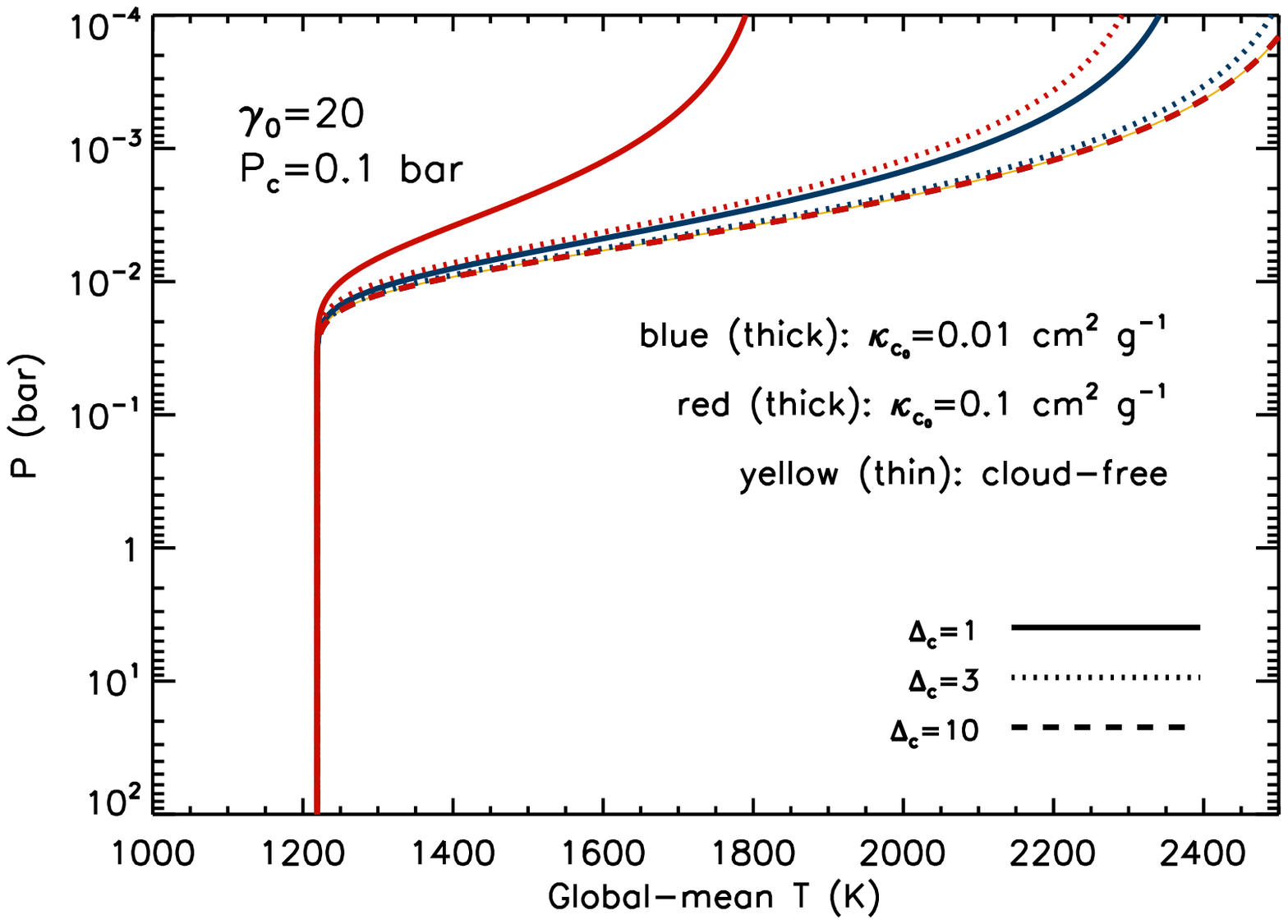}
\includegraphics[width=0.48\columnwidth]{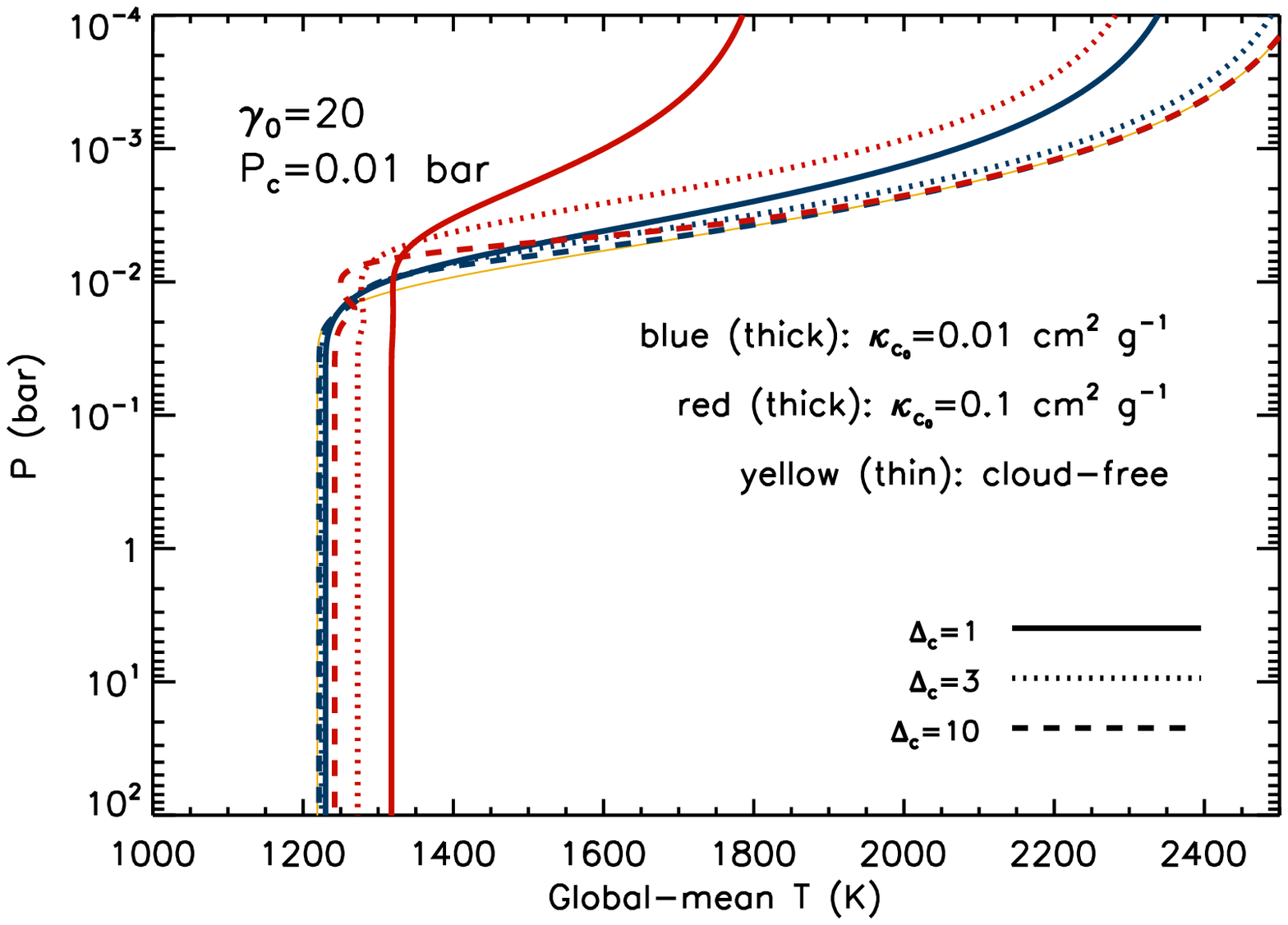}
\end{center}
\vspace{-0.2in}
\caption{Global-mean temperature-pressure profiles with purely absorbing Gaussian cloud/haze decks (assuming $\xi=1$), which illustrate their dual effects of cooling and warming the upper and lower atmosphere, respectively.  Shown are decks with three different thickness and two different opacity normalizations.  The left and right columns are for $P_{\rm c}=0.1$ bar and 0.01 bar, respectively.  The top, middle and bottom rows are for $\gamma_0=0.6, 2$ and 20, respectively.  In all of the panels, the cloud-free case is shown for comparison.}
\label{fig:cloud2}
\end{figure}

In Figure \ref{fig:cloud2}, we evaluate equation (\ref{eq:t4_global_cloud}) numerically to obtain the temperature-pressure profiles for models with various realizations of the idealized Gaussian cloud deck.  We retain the values of $T_{\rm eq}$, $g$, $\kappa_0$, $\kappa_{\rm S}$,  ${\cal A}$ and $\epsilon$ previously described in \S\ref{subsect:constant}.  Unlike in the case of a uniform cloud layer, the cooling/heating effect of the cloud deck depends on its location with respect to the shortwave photosphere as described in equation (\ref{eq:pressure_s}).  For $\kappa_{\rm S} = 0.006, 0.02$ and 0.2 cm$^2$ g$^{-1}$, we have $P_{\rm S} \approx 0.1$ bar, 30 mbar and 3 mbar, respectively.

In the top row of Figure \ref{fig:cloud2}, we adopt $\kappa_{\rm S} = 0.006$ cm$^2$ g$^{-1}$ such that $\gamma_0=0.6$.  In the top left panel, the cloud deck is essentially coincident with the shortwave photosphere.  Similar to the case of a uniform cloud layer, the cloud deck cools and warms the upper and lower atmosphere, respectively.  When the cloud deck is moved upwards to $P_{\rm c} = 0.01$ bar (top right panel of Figure \ref{fig:cloud2}), above the shortwave photosphere, the magnitude of the cooling effect remains somewhat the same but the warming effect becomes less pronounced.  In addition, the thinner cloud decks (e.g., $\Delta_{\rm c}=10$) imprint their Gaussian shapes onto the temperature-pressure profiles, causing a small temperature inversion.  

When the cloud deck is located far \emph{below} the shortwave photosphere, as shown in the bottom left panel of Figure \ref{fig:cloud2} where we adopt $\kappa_{\rm S} = 0.2$ cm$^2$ g$^{-1}$ such that $\gamma_0=20$, it still cools the upper atmosphere but has no effect on the lower atmosphere ($P \gtrsim 0.02$ bar).  Furthermore, only when the cloud deck is both thick (lower $\Delta_{\rm c}$) and dense (higher $\kappa_{\rm c_0}$) are the temperatures noticeably cooler compared to the cloud-free case.  When the cloud deck is shifted higher in altitude to $P_{\rm c}=0.01$ bar, the warming of the lower atmosphere now becomes noticeable, but is still more subdued than in the case when $P_{\rm c} < P_{\rm S}$.  

The middle row of Figure \ref{fig:cloud2} possesses the most physical richness compared to the other two rows.  The shortwave photosphere is now located at $P_{\rm S}\approx 30$ mbar, which corresponds to $\gamma_0=2$.  In the absence of a cloud deck, one would expect the temperature-pressure profiles to always possess temperature inversions.  For a Gaussian cloud deck, even the generalized condition in equation (\ref{eq:inversion_condition}) is insufficient to predict if a temperature inversion will occur.  When the cloud deck is tenuous ($\kappa_{\rm c_0}=0.01$ cm$^2$ g$^{-1}$), thin ($\Delta_{\rm c}=3,10$) and low-lying ($P_{\rm c}=0.1$ bar), temperature inversions occur, consistent with equation (\ref{eq:inversion_condition}).  However, if the cloud deck is made to be denser ($\kappa_{\rm c_0}=0.1$ cm$^2$ g$^{-1}$) it introduces a component to the profile where the temperature decreases with increasing altitude.  If it now sits higher in the atmosphere ($P_{\rm c}=0.01$ bar), then there are \emph{two} inverted components in the profile along with a non-inverted component.  By contrast, a low-lying ($P_{\rm c}=0.1$ bar), thick ($\Delta_{\rm c}=1$) and dense ($\kappa_{\rm c_0}=0.1$ cm$^2$ g$^{-1}$) cloud deck ``reverses" the $\gamma_0 > 1$ condition and produces a ``normal" temperature-pressure profile which behaves as if $\gamma_0<1$.  All of these features illustrate the complexity introduced by a simple, Gaussian cloud deck and demonstrate that there is no straightforward way to generalize the condition in equation (\ref{eq:inversion_condition}).

\subsection{Inclusion of shortwave scattering}

\begin{figure}
\begin{center}
\includegraphics[width=0.48\columnwidth]{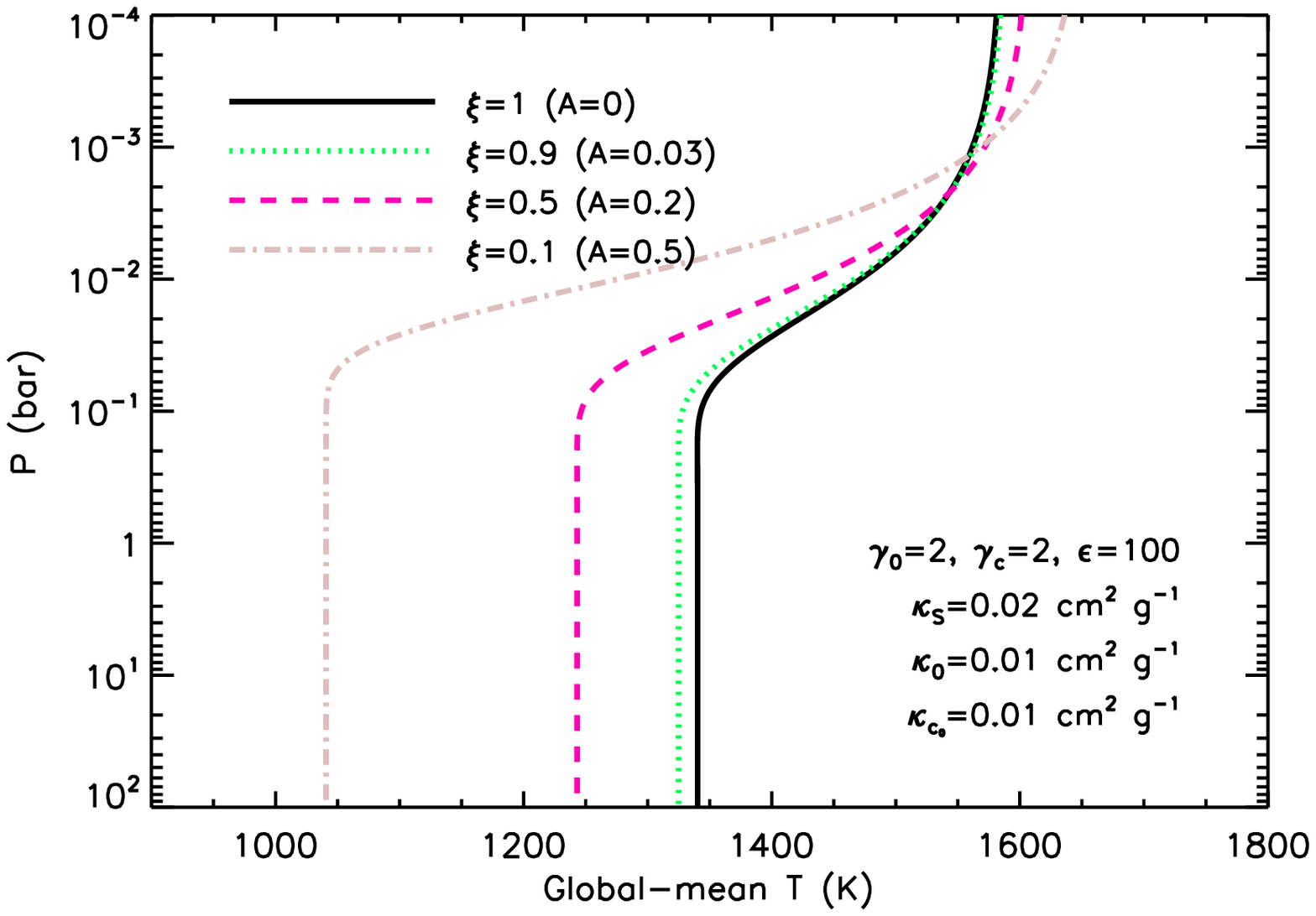}
\includegraphics[width=0.48\columnwidth]{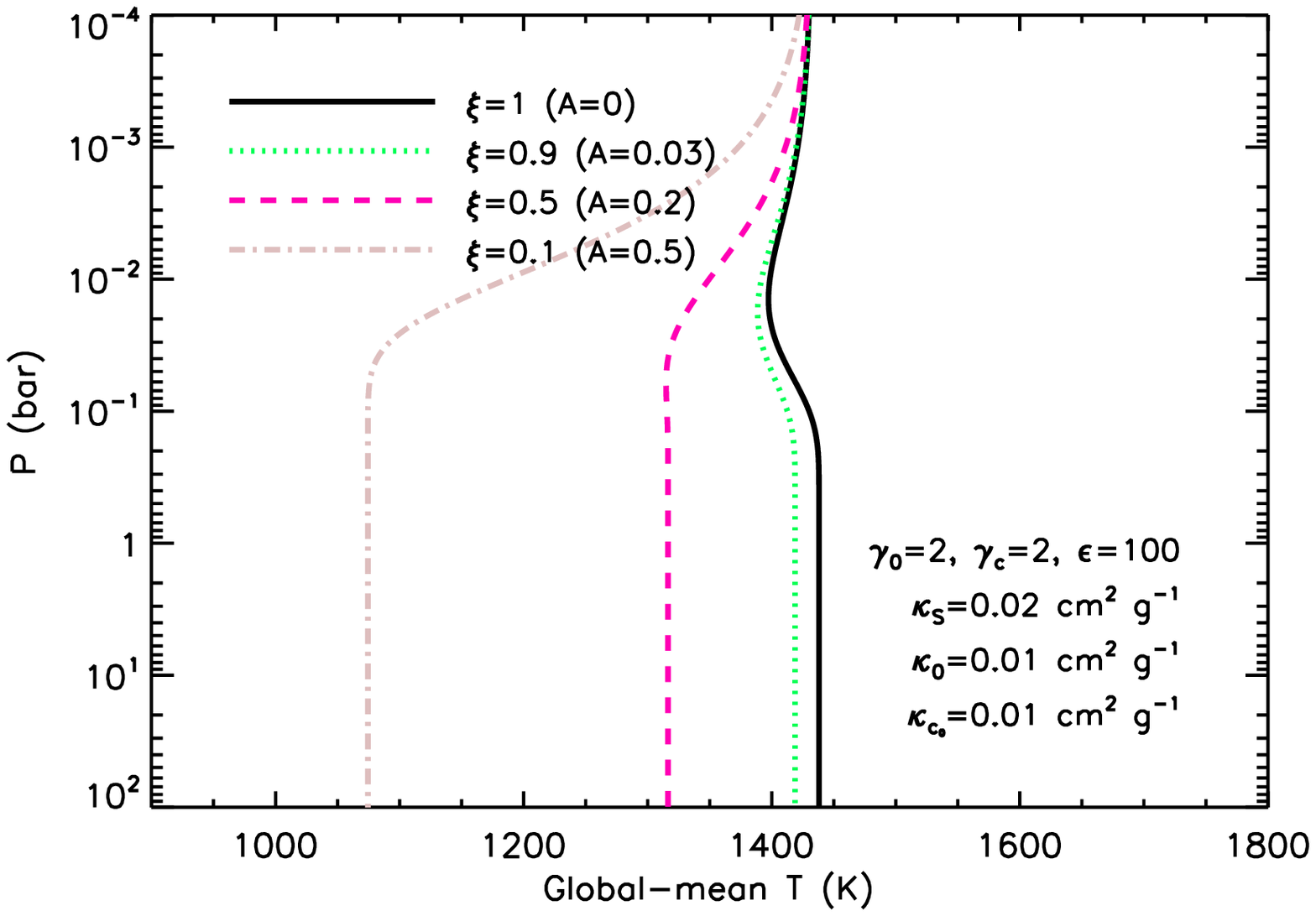}
\end{center}
\vspace{-0.2in}
\caption{Global-mean temperature-pressure profiles with uniform, scattering cloud/haze layers, which illustrate their dual effects of cooling and warming the lower and upper atmosphere, respectively.  The left and right panels are for $\kappa_{\rm c_0}=0$ and 0.01 cm$^2$ g$^{-1}$, respectively.  In both panels, the no scattering case ($\xi=1$) is shown for comparison.}
\label{fig:cloud_scat}
\end{figure}

We now allow for $\xi \ne 1$ and a non-zero Bond albedo (via equation [\ref{eq:albedo_xi}]) and investigate the effects of shortwave scattering by a uniform cloud/haze layer.  Setting $\kappa_{\rm c_0} = 0$ cm$^2$ g$^{-1}$, the left panel of Figure \ref{fig:cloud_scat} demonstrates that \textit{shortwave scattering has the opposite effect from that of an extra longwave opacity: it heats and cools the upper and lower atmosphere, respectively.}  This is further illustrated by comparing the left and right panels of Figure \ref{fig:cloud_scat}, where one sees that ``turning on" the cloud opacity $\kappa_{\rm c_0}$ partially negates the effects of scattering.  In general, scattering either produces or strengthens a temperature inversion as the shortwave photosphere is shifted to higher altitudes.

So far, we have used the terms ``cloud" and ``haze" synonymously.  We now attempt to better define the two terms.  By ``cloud", we refer to a layer of enhanced optical depth caused by certain chemical species condensing out of their gaseous forms.  In practice, these cloud decks will be situated roughly at the pressure level where their condensation curves intersect the temperature-pressure profile of the atmosphere.  Thus, clouds may be regarded as having a thermodynamic origin.  Such an approach of ``painting on" a cloud deck has been used to construct spectral-evolutionary models of brown dwarfs (e.g., \citealt{bht11}).  By contrast, a ``haze" layer may refer to an extra source of (absorption and scattering) opacity which is either uniform throughout the atmosphere  or is located at a pressure level which does not depend on a particular condensation curve.  Thus, hazes may be regarded as having a non-thermodynamic (e.g., photochemical) origin.  The simplicity of our models allows us to explore the effects of clouds and hazes on the temperature-pressure profile without having to perform a complicated coupling to a thermo-chemical model; the latter is required not only for self-consistency but also if predictions for chemical species and particle sizes are desired.

\section{Discussion}
\label{sect:discussion}

\subsection{Application of models to observations: HD 189733b}

We now make a tentative attempt to compare our models of the global-mean temperature-pressure profile to data points inferred from the published observations of the hot Jupiter HD 189733b.  We focus on HD 189733b as a case study because of the detection of haze in its atmosphere \citep{pont08,sing09,sing11}.  We make use of the transit spectroscopy, secondary eclipse and photometric phase curve observations of HD 189733b from the \textit{Hubble} and \textit{Spitzer Space Telescopes}.  In comparing one- to three-dimensional models, \cite{fortney10} demonstrated that the temperature-pressure profiles on the exoplanetary limb are good approximations to the true global-mean profile, making it natural to apply our formalism to temperatures derived from transit data.  

We extract the hemispherically-averaged brightness temperatures from the phase curves measured using \textit{Spitzer} at both 8 and 24 $\mu$m \citep{knutson07,knutson09}.  For the temperatures at the limb during transit, we use the average value, $(T_{\rm max}+T_{\rm min})/2$, as given in Table 3 of \cite{knutson09} for both the 8 and 24 $\mu$m temperatures, finding temperatures of 1135$\pm$52 K and 1102$\pm$67 K, respectively.  We also use the 3.6, 4.5, 5.8 and 16 $\mu$m secondary eclipse measurements of \cite{char08} and \cite{deming06} with the brightness temperatures given in \cite{ca11}.  As there is no phase curve information at these wavelengths, we applied a correction of one-half of the average day-night temperature contrast as measured at 8 and 24 $\mu$m (121$\pm$35 K) to estimate limb temperatures, finding temperatures of $1518 \pm 49$ K, $1197 \pm 57$ K, $1247 \pm 77$ K and $1217 \pm 62$ K at 3.6, 4.5, 5.8 and 16 $\mu$m, respectively.  The assumed pressures for these data points are model dependent and based upon the normalized contribution functions from \cite{knutson09}.  As such, the absolute pressure scale is uncertain and ultimately tied to the assumed (solar) model abundances and composition, with H$_2$O typically being the dominant opacity source in the near infrared.  Water has been identified in absorption in the \textit{Spitzer} emission spectra of \cite{grill08}.  

We augment the \textit{Spitzer} data with the temperatures associated with Rayleigh-scattering haze as detected and confirmed by \textit{Hubble Space Telescope (HST)} transmission spectra \citep{pont08,sing09,sing11}.  We use the temperature derived from the \textit{HST ACS} and \textit{NICMOS} data in \cite{sing09} of 1280$\pm$110 K.  We tie the pressure of the optical haze to that of the \textit{Spitzer} emission spectra measurements using the differences in altitude between the optical and near-infrared transit spectra as well as the difference from the transit geometry.  Denoting the radius of the hot Jupiter by $R_p$, the 0.75 $\mu$m optical transit radius is  $7\times 10^{-4} R_p/R_\star$ above the 8 $\mu$m transit radius of \cite{agol10}, which is approximately 1.7 scale heights at 1300 K or a difference in pressure of a factor of 5.5.  In addition, the slant-to-normal difference of the transit geometry decreases the pressure of the transit measurements by a factor of $\sqrt{2 \pi R_p/H} \approx 50$ at a given wavelength, with $H$ being the pressure scale height \citep{burrows01,fortney05,hansen08}.  Thus, the haze in the optical at 0.75 $\mu$m is a factor of about 275 lower in pressure than the 8 $\mu$m point from the emission spectrum, placing the haze at about the 0.7 mbar pressure level.  This estimated pressure value is in agreement with that of \cite{les08}, who assumed a haze composition of MgSiO$_3$.

While $P<0.1$ mbar data points do exist (Huitson, Sing et al., in preparation), it is important to note that our formalism is invalid at these low pressures because it does not include the physics of the thermosphere, including hydrodynamic escape (e.g., \citealt{mc09}).  It is likely that our assumptions of hydrostatic balance and radiative equilibrium break down significantly at these altitudes.  Therefore, we terminate the model comparisons for $P \le 0.1$ mbar.

The initial task is to cut down on the number of parameters used in our models.  We estimate that $T_{\rm irr} \approx 1697$ K $(1-{\cal A})^{1/4}$ \citep{bouchy05} and $g \approx 21.88$ m s$^{-2}$ \citep{torres08}.  Since the data points are for $P < 1$ bar, they set no constraint on the temperatures at depth and thus are unaffected by the value of $T_{\rm int}$ adopted --- for simplicity, we set $T_{\rm int}=0$ K.  Again for simplicity, we set $\kappa_{\rm c_0}=0$ cm$^2$ g$^{-1}$, namely that any extra longwave absorption by the cloud/haze present may be assimilated into the longwave opacity function $\kappa_{\rm L}$.  Therefore, we are left with a 4-parameter model to be compared to 7 data points.  The assumption of a uniform scattering opacity is not unreasonable given the fact that \cite{sing11} detect a transmission spectrum of HD 189733b consistent with a haze layer, producing Rayleigh scattering, at $\sim 1$ mbar, spanning about 8 pressure scale heights (albeit at somewhat higher altitudes than where we are making our model comparisons).  

\begin{figure}
\begin{center}
\includegraphics[width=0.48\columnwidth]{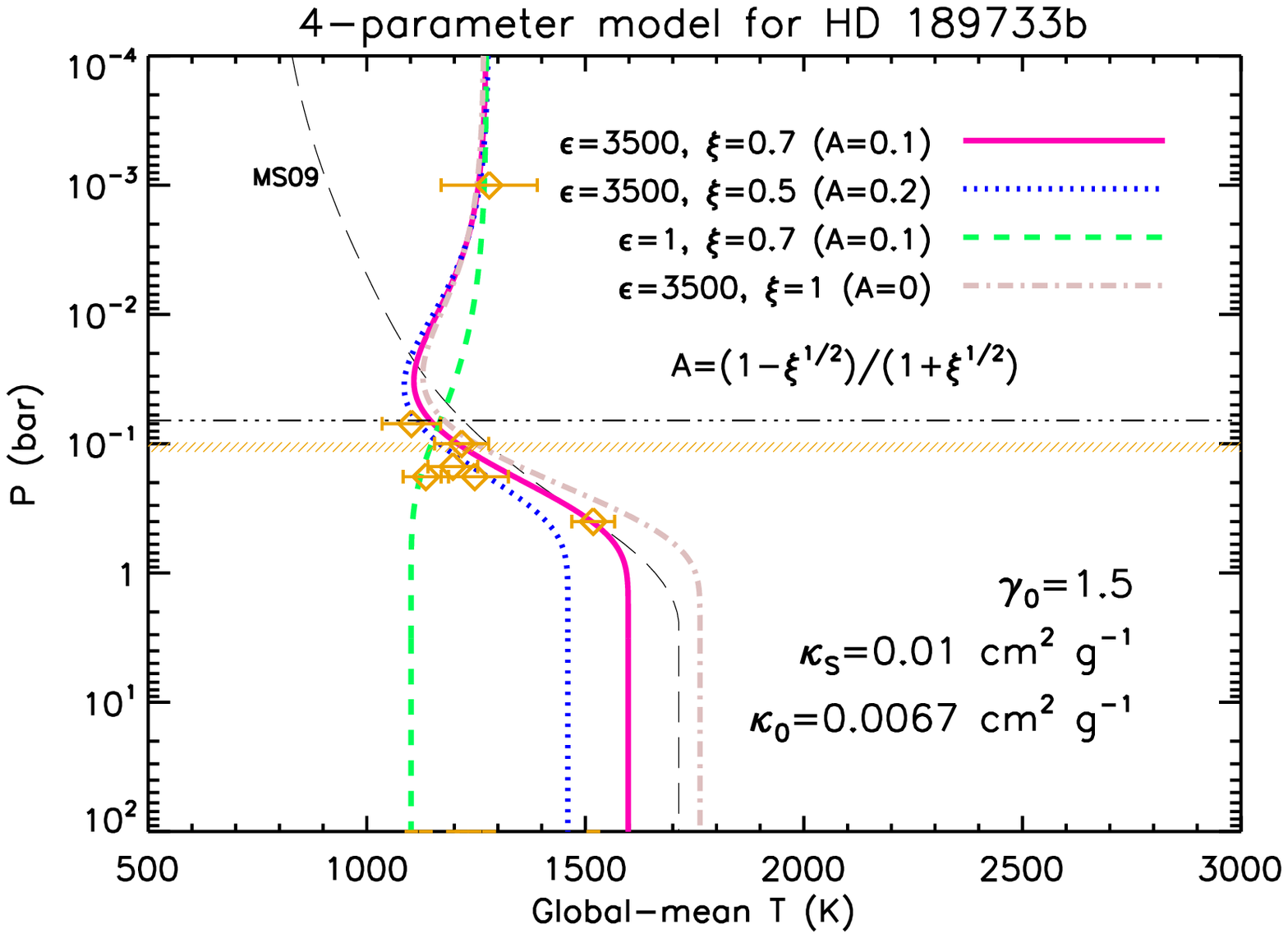}
\includegraphics[width=0.48\columnwidth]{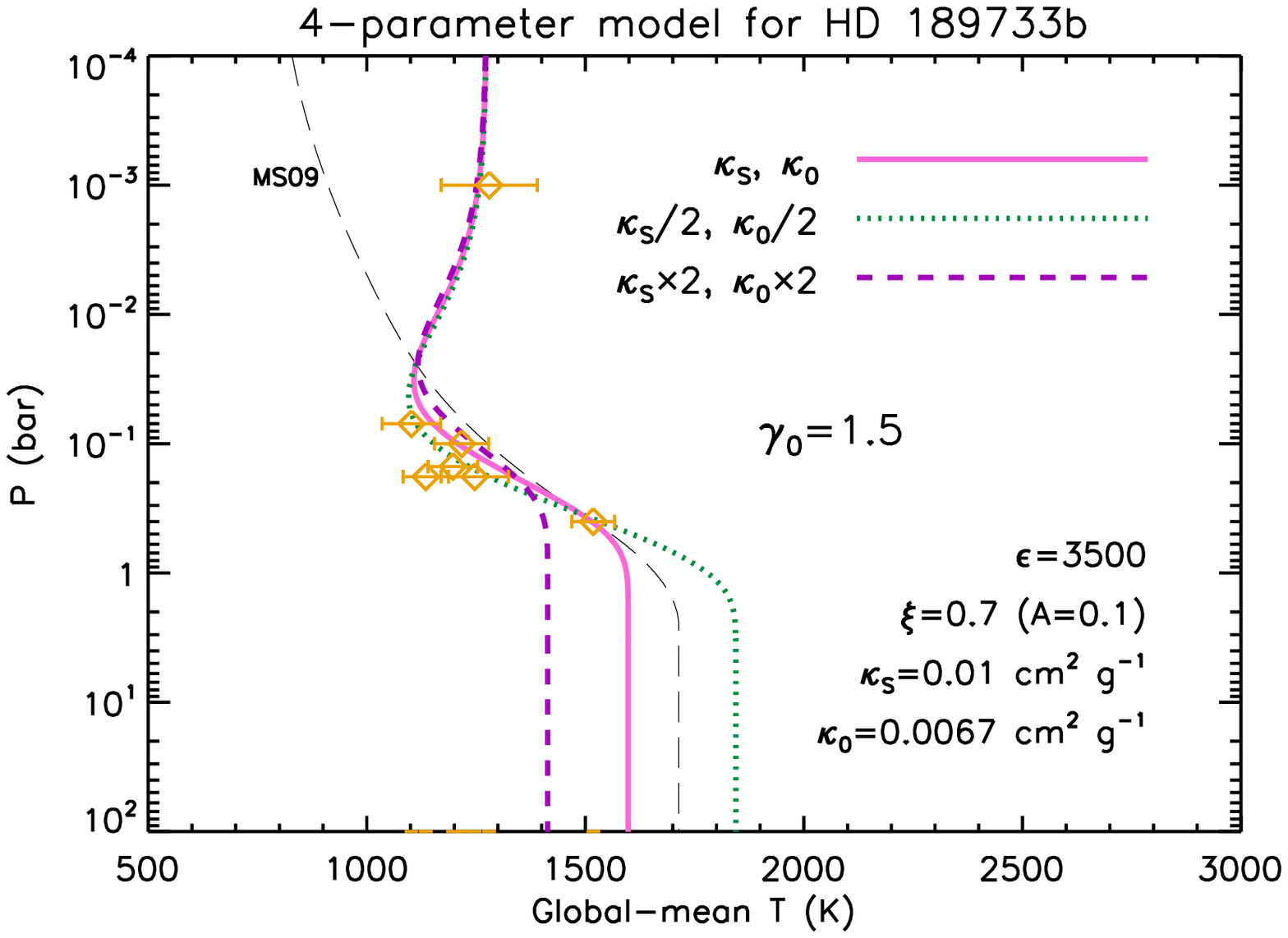}
\end{center}
\vspace{-0.2in}
\caption{Comparison of global-mean temperature-pressure profiles to data points inferred from observations of HD 189733b (see text for details and caveats).  The observed system parameters allow us to set $T_{\rm irr}=1697$ K, while we set $T_{\rm int}=0$ K and $\kappa_{\rm c_0}=0$ cm$^2$ g$^{-1}$ for simplicity. The left panel fixes the shortwave opacity and longwave opacity normalization and illustrates the effects due to the variation of the other two parameters.  The right panel fixes $\epsilon$ and $\xi$ and varies the opacities.  In the left panel, the horizontal, dash-triple-dot line shows the approximate position of the longwave photosphere, while the hatched, yellow region indicates the approximate location of the shortwave photon deposition depth (corresponding to the pair of $\xi$ values adopted).  We terminate all of the models at $P=0.1$ mbar as our formalism does not treat the physics of the thermosphere.  For comparison, we include the temperature-pressure profile from Madhusudhan \& Seager (2009), labelled ``MS09", which was obtained via an abundance and temperature retrieval method.}
\label{fig:cloud_fit}
\end{figure}

We begin with a $\gamma_0=1.5$ model with some scattering present ($\xi=0.7, {\cal A} \approx 0.1$) and without including the effect of collision-induced absorption ($\epsilon=1$), using equation (\ref{eq:t4_global_cloud0}), as shown in the left panel of Figure \ref{fig:cloud_fit}.  It is clear that while such a model is roughly consistent with the increase in temperature from $\sim 0.1$ bar to $\sim 1$ mbar, it fails to produce the increase in temperature with pressure at $P \gtrsim 0.1$ bar.  By contrast, a model with no scattering ($\xi=1, {\cal A}=0$) and which includes collision-induced absorption ($\epsilon=3500$) is roughly consistent with both of these trends, but does not quite pass through the data points.  As we retain $\epsilon=3500$ and allow scattering to be present ($\xi=0.7, {\cal A}\approx 0.1$), the lower atmosphere becomes cooler, thus allowing the model to be consistent with the data points.  Too much scattering ($\xi=0.5, {\cal A} \approx 0.2$) results in a temperature-pressure profile which is cooler than indicated by the data points.

It is reasonable to ask if our model comparisons are sensitive to the assumed values of the opacities.  In the right panel of Figure \ref{fig:cloud_fit}, we retain $\gamma_0=1.5$ but calculate two more models where we reduce and increase both of the shortwave and longwave opacities by a constant factor of 2.  It is clear that even with $\gamma_0$ kept constant, the ``turn" in the temperature-pressure profile --- i.e., first decreasing, then increasing, temperature with increasing altitude --- is sensitive to the values of the opacities.  Lower opacities result in the ``turn" residing at higher pressures.  The temperatures at depth ($P \gtrsim 1$ bar) vary sensitively when the opacities are varied due to the changing depths at which the stellar photon energy is being deposited (see equation [\ref{eq:deposition}]).

For comparison, we include in Figure \ref{fig:cloud_fit} the temperature-pressure profile for HD 189733b from \cite{madhu09}, which was obtained via an abundance and temperature retrieval method applied to infrared transit observations from both the \textit{HST} and the \textit{Spitzer Space Telescope}.  Given the uncertainties associated with both theory and observation, the profile is broadly consistent with our models for $P \gtrsim 0.1$ bar but does not include the temperature inversion from $\sim 0.1$ bar to $\sim 1$ mbar.  This is unsurprising because the data point at about 1 mbar was not included in the original analysis of \cite{madhu09}.

One may find the status of HD 189733b as the prototypical example of an inversion-less hot Jupiter to be at odds with the increase in temperature from $\sim 0.1$ bar to $\sim 1$ mbar and our interpretation of it using a $\gamma_0 > 1$ model.  This dilemma is resolved if one realizes that there are two types of temperature inversions: deep- versus meso-atmospheric inversions.  The former refers to temperature inversions probed by near infrared water lines, typically at $\sim 0.1$--1 bar: at these pressure levels, there is general consensus that HD 189733b and HD 209458b are the prototypes of hot Jupiters without and with temperature inversions in their atmospheres, respectively \citep{burrows07,char08,fortney08,madhu09}.  The latter refers to temperature inversions from $\sim 0.1$ bar to $\sim 1$ mbar, as probed by the \textit{HST} transmission spectra of \cite{pont08} and \cite{sing09,sing11}.  

\textit{Collectively, our models make a crude prediction for the Bond albedo of HD 189733b: ${\cal A} \approx 0.1$.}  The Bond albedo is not an observable quantity; rather, it is the geometric albedo which is measured, as has been done for HD 209458b by \cite{rowe08}.  (See \citealt{burrows08} for the implications of albedo measurements on modelling.)  The conversion from the geometric to the Bond albedo requires integrating over both frequency and phase angle.  For example, \cite{bs09} estimate that the \textit{spherical albedo}\footnote{The Bond albedo is obtained by integrating the spherical albedo over frequency.} ${\cal A}_s$ and the geometric albedo ${\cal A}_g$ are related by ${\cal A}_g \approx 0.8 {\cal A}_s$ in their models of exoplanetary atmospheres with Rayleigh scattering.  The reader is again referred to \S3.4 of \cite{seager10} for a detailed discussion of exoplanetary albedos.

\subsection{Semi-analytical models as a guide for three-dimensional simulations}

The ability of our models to predict the temperature at depth ($T_\infty$) for a given set of parameters provides a useful guide to three-dimensional simulations of atmospheric circulation which utilize dual-band, two-stream radiative transfer \citep{hfp11}.  Such simulations require the specification of the initial temperature and velocity fields.  The simplest assumption is to initiate them from a state of windless isothermality, which then requires the specification of a constant, initial temperature $T_{\rm init}$.  Computational efficiency is optimized when the initial temperature field is as close to radiative equilibrium as possible.  Physically, we expect the temperature to approach radiative equilibrium at depth ($P \gtrsim 10$ bar).  Therefore, selecting $T_{\rm init}=T_\infty$ is a good choice.  As the simulation proceeds, the upper atmosphere ($P \lesssim 10$ bar) adjusts itself to temperatures consistent with dynamical-radiative equilibrium, while the lower atmosphere ($P \gtrsim 10$ bar) remains in radiative equilibrium.  A state of quasi-equilibrium is reached when there is dynamical-radiative equilibrium within the upper atmosphere, while the lower atmosphere ceases to transfer significant amounts of energy upwards.  Semi-analytical models such as the ones presented in this study alleviate the need to perform a tedious parameter search for the appropriate value of $T_{\rm init}$ to adopt and therefore increase the efficiency of utilizing general circulation models to study the atmospheres of hot Jupiters.

\subsection{Future work}

A useful extension of our models will be to consider the dependence of the radius, at a given wavelength, of a hot Jupiter on various quantities,
\begin{equation}
R = R\left( T_\infty, T_{\rm int}, g \right).
\label{eq:radius}
\end{equation}
The temperature at depth $T_\infty$ is a reasonable proxy for the atmospheric effects due to the variation of absorption and scattering.  Our formalism can then be used to estimate a value for $T_\infty$, which will provide a lower boundary condition for models of hot Jovian interiors.  While the elucidation of equation (\ref{eq:radius}) is beyond the scope of the present study, such an exercise will be useful for interpreting the inflated radii of some hot Jupiters (e.g., \citealt{laughlin11}).  Specifically, combining our atmospheric models with those of hot Jovian interiors will allow a connection to the observed (transit) radius, thereby allowing us to rule out models which over-inflate a given hot Jupiter.

\subsection{Summary}

The salient points of our study may be summarized as follows:
\begin{itemize}

\item We have presented a semi-analytical model for the global-mean temperature-pressure profile of a hot Jovian atmosphere which requires the specification of 7 parameters: the intrinsic heat flux (via $T_{\rm int}$), the flux of stellar irradiation (via $T_{\rm irr}$ or $T_{\rm eq}$), the surface gravity of the exoplanet ($g$), the strength of shortwave scattering ($\xi$ or via ${\cal A}$), the shortwave opacity ($\kappa_{\rm S}$) and the longwave opacity (via $\kappa_0$ and $\epsilon$).  Allowing for an extra source of longwave absortion due to the presence of a uniform cloud/haze layer adds another parameter ($\kappa_{\rm c_0}$), while a purely absorbing Gaussian cloud/haze deck adds another 2 parameters to the system ($\kappa_{\rm c_0}$, $\Delta_{\rm c}$ and $P_{\rm c}$; but sets $\xi=1$).

\item If a cloud/haze layer contributes an extra source of longwave opacity, then the main effect is to cool and heat the upper and lower atmosphere, respectively, analogous to the greenhouse effect on Earth.

\item If a cloud/haze layer causes scattering in the shortwave, then the main effect is to heat and cool the upper and lower atmosphere, respectively, thus counteracting its absorbing effect in the longwave --- the anti-greenhouse effect.  Scattering shifts both the shortwave photosphere and the photon deposition depth to higher altitudes.

\item Besides their utility in developing physical intuition, our semi-analytical models provide a guide to three-dimensional simulations of atmospheric circulation.  In particular, they pin down the temperature at depth for a given set of parameters, which serves as the initial condition for the temperature field.

\item We have made a tentative comparison of 4-parameter models to the temperature-pressure data points inferred from the \textit{Hubble} and \textit{Spitzer} observations of HD 189733b and estimate that its Bond albedo is approximately 0.1.  Future observations of HD 189733b will corroborate or refute some of our models. 

\item The simplicity and versatility of our semi-analytical models allow for an easy comparison to observations.  Observers wishing to use our models should refer to equation (\ref{eq:t4_global_cloud0}) and set $T_{\rm int}=0$ K and $\gamma^{-1}_{\rm c}=0$ as a starting point, thereby distilling the model down to having only 4 parameters.

\end{itemize}

\begin{table*}
\centering
\caption{Table of commonly used symbols}
\label{tab:symbols}
\begin{tabular}{lcc}
\hline\hline
\multicolumn{1}{c}{Symbol} & \multicolumn{1}{c}{Meaning} & \multicolumn{1}{c}{Units}\\
\hline
\vspace{2pt}
$\mu = \cos\theta$ & cosine of latitude & --- \\
$m$ & column mass per unit area of atmosphere & g cm$^{-2}$ \\
$m_0$ & column mass per unit area at bottom of model atmosphere & g cm$^{-2}$ \\
$P$ & vertical pressure & bar or dyne cm$^{-2}$ \\
$P_0$ & pressure at bottom of model atmosphere & bar or dyne cm$^{-2}$ \\
$P_{\rm c}$ & pressure level where cloud/haze deck is located & bar or dyne cm$^{-2}$ \\
$\xi$ & shortwave scattering parameter (ratio of absorption to total opacity) & --- \\
$\epsilon$ & factor for enhancement of longwave optical depth (at $P=P_0$) due to collision-induced absorption & --- \\
$\kappa_{\rm S}$ & shortwave opacity & cm$^2$ g$^{-1}$ \\
$\kappa_{\rm L}$ & longwave opacity & cm$^2$ g$^{-1}$ \\
$\kappa_0$ & longwave opacity normalization & cm$^2$ g$^{-1}$ \\
$\kappa_{\rm c}$ & additional longwave opacity due to cloud/haze deck & cm$^2$ g$^{-1}$ \\
$\kappa_{\rm c_0}$ & longwave opacity normalization of cloud/haze deck & cm$^2$ g$^{-1}$ \\
$\gamma \equiv \kappa_{\rm S}/\kappa_{\rm L}$ & ratio of shortwave to longwave opacities & --- \\
$\gamma_0 \equiv \kappa_{\rm S}/\kappa_0$ & ratio of shortwave opacity to longwave opacity normalization & --- \\
$\gamma_{\rm c} \equiv \kappa_{\rm S}/\kappa_{\rm c_0}$ & ratio of shortwave opacity to longwave opacity normalization due to cloud/haze & --- \\
$\tau_{\rm L} = \int \kappa_{\rm L} ~dm$ & longwave optical depth & --- \\
$\tau_{\rm S} = \kappa_{\rm S} m/\xi$ & shortwave optical depth & --- \\
$\tau_0 \equiv \kappa_0 m_0$ & longwave optical depth normalization in absence of collision-induced absorption & --- \\
$\tau \equiv \kappa_{\rm L} m$ & --- & --- \\
$\tau_{\rm c}$ & additional longwave optical depth due to cloud/haze deck & --- \\
$\Delta_{\rm c}$ & thickness parameter of cloud/haze deck & --- \\
$T_{\rm int}$ & blackbody-equivalent temperature associated with internal heat flux & K \\
$T_{\rm irr}$ & irradiation temperature & K \\
$T_{\rm eq} = T_{\rm irr}/\sqrt{2}$ & equilibrium temperature of hot Jupiter & K \\
$\bar{T}$ & global-mean temperature of atmosphere & K \\
$T_\infty$ & temperature at depth & K \\
\hline
\hline
\end{tabular}\\
Note: the term ``opacity" refers only to absorption opacities.
\end{table*}

\vspace{0.2in}
\noindent
\textit{K.H. acknowledges support from the Zwicky Prize Fellowship and the Star and Planet Formation Group at ETH Z\"{u}rich, and benefited from stimulating interactions with the Astrophysics Group at Exeter University.  W.H. acknowledges support by the European Research Council under the European Community's 7th Framework Programme (FP7/2007--2013 Grant Agreement no. 247060).  F.P. acknowledges support by a STFC Advanced Fellowship.  We thank Isabelle Baraffe, Tristan Guillot, Esther Buenzli, Adam Burrows, Hans Martin Schmid, Sascha Quanz, Giovanna Tinetti and Nikku Madhusudhan for useful conversations.}

\appendix

\section{Notes on collimated beam and two-stream approximations}
\label{append:two_stream}

In the collimated beam approximation, the shortwave intensity may be written as
\begin{equation}
I_{\rm S}\left(\mu^\prime\right) = I_+ ~\delta\left( \mu^\prime - \mu \right) + I_- ~\delta\left( \mu^\prime + \mu \right),
\label{eq:collimated_s}
\end{equation}
such that its moments are
\begin{equation}
\begin{split}
&J_{\rm S} = \frac{1}{2} \int^{1}_{-1} ~I_{\rm S} ~d\mu^\prime = \frac{1}{2} \left( I_+ + I_- \right), \\
&H_{\rm S} = \frac{1}{2} \int^{1}_{-1} ~\mu^\prime I_{\rm S} ~d\mu^\prime = \frac{\mu}{2} \left( I_+ - I_- \right), \\
&K_{\rm S} = \frac{1}{2} \int^{1}_{-1} ~\mu^{\prime 2} I_{\rm S} ~d\mu^\prime = \frac{\mu^2}{2} \left( I_+ + I_- \right), \\
\end{split}
\label{eq:two_stream_moments}
\end{equation}
where $I_+=I_+(m)$ and $I_-=I_-(m)$ denote the outgoing and incoming shortwave intensities, respectively.  The cosine of the angle between the beam and the vertical axis is represented by $\mu$, while the Dirac delta function(al), $\delta(\mu^\prime \pm \mu)$, describes the sharply peaked angular distributions.  A direct consequence of equation (\ref{eq:two_stream_moments}) is that $K_{\rm S}/J_{\rm S} = \mu^2$ as stated in equation (\ref{eq:eddington2}).  Using equations (\ref{eq:h_s0}) and (\ref{eq:two_stream_moments}), one may then derive the expression relating the Bond albedo and $\xi$, as stated in equation (\ref{eq:albedo_xi}), by recognizing that
\begin{equation}
{\cal A} = \frac{I_+}{I_-}.
\end{equation}

For completeness, we examine the longwave intensity in the two-stream approximation by assuming
\begin{equation}
I_{\rm L}\left( \mu^\prime \right) = \alpha + \beta \mu^\prime.
\label{eq:two_stream}
\end{equation}
Equation (\ref{eq:two_stream}) describes a nearly isotropic radiation field, but in this approximation it is sufficient to describe it by two rays.  Taking moments of equation (\ref{eq:two_stream}), we get
\begin{equation}
{\cal E}_1 \equiv \frac{K_{\rm L}}{J_{\rm L}} = \frac{1}{3}, ~{\cal E}_2 \equiv \frac{H_{\rm L}}{J_{\rm L}} = \frac{\beta}{3\alpha}.
\end{equation}
We next derive the $\mu$-values of the two rays in a heuristic fashion; more general derivations exist (e.g., \citealt{toon89}), but we are merely seeking to gain physical intuition for the value of the second Eddington coefficient.  At the top of the atmosphere, the incoming intensity is zero since we are assuming that the star radiates negligibly at long wavelengths.  We sample the radiation field (equation [\ref{eq:two_stream}]) with two rays, similar to equation (\ref{eq:collimated_s}), and denote the outgoing longwave intensity by $I_0$:
\begin{equation}
J_{\rm L} = \frac{I_0}{2}, ~H_{\rm L} = \frac{\mu I_0}{2}, ~K_{\rm L} = \frac{\mu^2I_0}{2},
\end{equation}
whence
\begin{equation}
{\cal E}_1 = \mu^2, ~{\cal E}_2 = \mu.
\end{equation}
By requiring that ${\cal E}_1=1/3$ be satisfied, we obtain
\begin{equation}
\mu^2 = \frac{1}{3} \implies {\cal E}_2 = \frac{1}{\sqrt{3}} \approx 0.58.
\end{equation}
The term ``two-stream" follows from the fact that only two rays at $\mu = \pm 1/\sqrt{3}$ are needed in this approximation.

\section{Notes on the photon deposition depth}
\label{append:deposition}

Using equations (\ref{eq:shortwave3}) and (\ref{eq:h_meaning}), the first moment of the shortwave intensity may be rewritten as
\begin{equation}
H_{\rm S} = H_0 ~\mu \exp{\left(- \frac{\sqrt{\xi} \tau_{\rm S}}{\mu} \right)},
\end{equation}
where $H_0 \equiv -\sigma_{\rm SB}T^4_{\rm irr}/4\pi$.  Taking the latitudinal average, we get
\begin{equation}
\langle H_{\rm S} \rangle = 2 H_0 ~E_3\left( \sqrt{\xi} \tau_{\rm S} \right).
\end{equation}
When $\sqrt{\xi} \tau_{\rm S}=0$, we have $\langle H_{\rm S}(0) \rangle = H_0$.  We define the photon deposition depth as the layer at which the incident stellar flux diminishes to $e^{-1} \approx 0.368$ of its initial value, which is described by
\begin{equation}
\frac{\langle H_{\rm S} \rangle}{\langle H_{\rm S}\left(0\right) \rangle} = 2 E_3\left( \sqrt{\xi} \tau_{\rm S} \right) \approx 0.368.
\label{eq:deposit_eqn}
\end{equation}
If we write the photon deposition optical depth as occurring at $\sqrt{\xi} \tau_{\rm S} = \varpi$, then solving equation (\ref{eq:deposit_eqn}) numerically yields $\varpi \approx 0.63$.

\section{An identity involving the derivative of $\tilde{Q}$}
\label{append:identity}

For an arbitrary function ${\cal X} = {\cal X}(x,y)$, we have
\begin{equation}
\frac{\partial}{\partial x} \int^{y_2\left(x\right)}_{y_1\left(x\right)} ~{\cal X}\left(x,y\right) ~dy = \int^{y_2\left(x\right)}_{y_1\left(x\right)} ~\frac{\partial {\cal X}\left(x,y\right)}{\partial x} ~dy+ \frac{d y_2\left(x\right)}{dx} {\cal X}\left(x,y_2\right) - \frac{d y_1\left(x\right)}{dx} {\cal X}\left(x,y_1\right).
\end{equation}
We now apply the preceding equation to
\begin{equation}
\frac{\partial \tilde{Q}\left(m,\mu,\phi\right)}{\partial m} = \frac{\partial}{\partial m} \int^{m_2\left(m\right)}_{m_1\left(m\right)} ~Q\left(m^\prime,\mu,\phi\right) ~dm^\prime.
\end{equation}
Since $Q = Q(m^\prime,\mu,\phi)$ only, we have $\partial Q(m^\prime,\mu,\phi)/\partial m=0$.  For a constant $m_2 \rightarrow \infty$, we have
\begin{equation}
\lim_{m_2\rightarrow \infty} \frac{dm_2}{dm} ~Q\left(m_2,\mu,\phi\right) = 0.
\end{equation}
The only non-zero term which remains is
\begin{equation}
-\frac{d m_1}{dm} Q\left(m_1,\mu,\phi\right) = -Q\left(m,\mu,\phi\right),
\end{equation}
where we have used $m_1 = m$.  We thus prove the identity in equation (\ref{eq:q_heat2}).

\section{Notes on mathematical operators}
\label{append:operators}

As there is some ambiguity in the astrophysical/astronomical community regarding the use of mathematical operators, it is useful to clarify the meaning of the operators used in this study.  There is generally no confusion over the ``=" sign, but when it is used to equate an algebraic expression to a numerical answer, it means that \emph{no} approximations (including rounding-off) are made, i.e., the answer is \emph{exact}.  If either an approximation is taken or rounding-off is performed, then the ``$\approx$" sign is employed.  The much-abused ``$\sim$" sign means ``on the order of"; it does \emph{not} mean ``is proportional to" (``$\propto$").  The ``$\rightarrow$" sign means ``tends to", i.e., the asymptotic value of a quantity.  Finally, the ``$\equiv$" sign means ``is defined as".


\label{lastpage}

\end{document}